INSTITUTO TECNOLÓGICO AUTÓNOMO DE MÉXICO

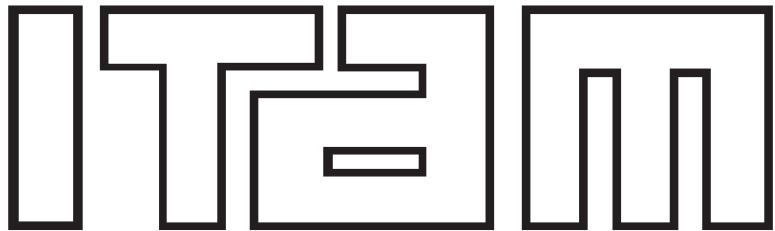

ESTIMACIÓN DE LA INICIAL DE REFERENCIA
UTILIZANDO SIMULACIÓN

INVESTIGACIÓN DOCUMENTAL
(ÁREA DE CONCENTRACIÓN: LOGÍSTICA Y PRODUCCIÓN)

# T E S I N A

QUE PARA OBTENER EL TÍTULO DE
**INGENIERO INDUSTRIAL**
P R E S E N T A
EMILIANO DÍAZ SALAS-PORRAS

MÉXICO, D.F.      2011

Con fundamento en los artículos 21 y 27 de la Ley Federal del Derecho de Autor y como titular de los derechos moral y patrimonial de la obra titulada "ESTIMACIÓN DE LA INICIAL DE REFERENCIA UTILIZANDO SIMULACIÓN", otorgo de manera gratuita y permanente al Instituto Tecnológico Autónomo de México y a la Biblioteca Raúl Bailléres Jr., Autorización para que fijen la obra en cualquier medio, incluido el electrónico, y la divulguen entre sus usuarios, profesores, estudiantes o terceras personas, sin que pueda percibir por tal divulgación una contraprestación.

<p align="center">Emiliano Díaz Salas-Porras</p>

<p align="center">______________________<br>
Fecha</p>

<p align="center">______________________<br>
Firma</p>

# ESTIMACIÓN DE LA INICIAL DE REFERENCIA UTILIZANDO SIMULACIÓN

**Emiliano Díaz Salas-Porras**

## Resumen

Se analizo el método para aproximar, por simulación, la inicial de referencia propuesta por Bernardo y Smith [2000] con el objetivo de mejorar el procedimiento para obtener estimadores consistentes, y para permitir la estimación del error por medio de intervalos asintóticos. En este sentido se derivo la Varianza del estimador de Bernardo y con esta se construyeron intervalos de probabilidad que permiten relacionar el tamaño de la muestra simulada con el error de estimación. Además, se investigó el uso de una técnica de reducción de varianza (números aleatorios comunes) para obtener estimaciones más precisas con menores tamaños de muestra y se encontró que se reduce considerablemente el error de estimación en algunos de los ejemplos estudiados y en otros casos el error es nulo ya que el estimador resulta no depender de la muestra.

**Palabras clave:** Estadística Bayesiana, Inicial de Referencia, Números Aleatorios Comnes

# Índice

# Índice de Figuras

## Índice de Tablas

# 1. INTRODUCCIÓN

## 1.1 Antecedentes

### 1.1.1 Estimación Bayesiana

La estadística frecuentista se conforma por una serie de métodos y soluciones a problemas concretos propuestos a menudo de forma intuitiva y no en un cuerpo de teoría coherente. Por ejemplo, es común encontrar en la literatura de estadística frecuentista que se proponga el método de máxima verosimilitud ó el método de momentos [Mood et al., 1974, p.273] como posibles soluciones al problema de estimación puntual. Inclusive existen criterios, como el sesgo y el error cuadrado medio [Mood et al., 1974, p.288] para evaluar el desempeño de los estimadores que resultan de estos respectivos métodos. Sin embargo, no existe un planteamiento o justificación formal de porque estos métodos deben ser usados para resolver el problema. La estadística Bayesiana, en cambio, es un cuerpo de teoría, construido sistemáticamente para resolver el problema de decisiones tomadas en ambiente de incertidumbre. El problema de estimación parametral se trata en este contexto como una decisión en ambiente de incertidumbre.

En la teoría de decisiones un problema de decisión bajo incertidumbre consiste en la cuarteta $\{D, E, C, \prec\}$ donde:

- $D$ es un conjunto exhaustivo de opciones
- $E$ es el conjunto de eventos inciertos relevantes
- $C$ es el conjunto de consecuencias pertinentes
- $\prec$ es una relación de orden sobre el conjunto $C$ que indica la preferencia que se tiene entre cualesquiera dos consecuencias

Resolver un problema de decisión equivale a escoger un elemento del conjunto $D$. La teoría Bayesiana, partiendo de unos *Axiomas de Coherencia* [Bernardo y Smith, 2000] indica que la única forma libre de errores sistemáticos de tomar una decisión en ambiente de incertidumbre es midiendo cualquier incertidumbre relevante al problema de decisión por medio de una función de probabilidad subjetiva y ordenando las consecuencias del problema de decisión mediante una función de pérdida. La función de probabilidad y la función de pérdida son medidas subjetivas de incertidumbre y de preferencia, respectivamente: son las de un tomador de decisiones específico. Es importante tomar en cuenta que, en la teoría Bayesiana, cualquier tipo de incertidumbre se debe medir con una función de probabilidad y no únicamente fenómenos que presentan variabilidad como ocurre en la teoría de probabilidad clásica. Por ejemplo, si existe un parámetro relevante al problema de decisión cuyo valor es fijo pero desconocido, la incertidumbre que se tiene acerca del valor verdadero de este parámetro se debe medir con una función de probabilidad. Conforme a esta teoría, la decisión óptima para un problema es aquella que minimiza la esperanza con respecto a la función de probabilidad escogida, de la función de pérdida. Otro aspecto importante de la teoría Bayesiana es que los *Axiomas de Coherencia* implican los *Axiomas de Kolmogorov* (que bajo esta teoría no son axiomas sino resultados). Por esta razón, la función de probabilidad que se use para medir la incertidumbre debe cumplir todas las propiedades de la probabilidad clásica incluyendo el Teorema de Bayes que en este contexto cobra una relevancia especial. Como se discute a continuación, el problema de inferencia paramétrica se puede resolver con la metodología Bayesiana.

Sean $F_1, F_2, ..., F_n$ eventos mutuamente exclusivos y $G$ otro evento tal que $P(G) \neq 0$. . Entonces, según el Teorema de Bayes [ver e.g., Chung, 1972]:

$$P(F_j \mid G) = \frac{P(F_j, G)}{P(G)} = \frac{P(G \mid F_j)P(F_j)}{\sum_{i=1}^{n} P(G \mid F_i)P(F_i)}$$

Si se interpreta a $F_j$ como una hipótesis entre *n* posibles, una de las cuales es cierta y se tiene una medida de la incertidumbre inicial acerca de la veracidad de cada una en $P(F_j)$ para $j = 1,...,n$, entonces el teorema muestra como actualizar la medida de probabilidad subjetiva a la luz de nueva evidencia (la ocurrencia del evento *G*) siempre y cuando contemos con $P(G \mid F_j)$ una medida de la probabilidad de que sea cierto el evento *G* dado que la hipótesis $F_j$ es cierta.

En esta investigación interesa más el Teorema de Bayes para densidades de probabilidad continuas y su interpretación Bayesiana. Según este teorema, si *X* y *Y* son variables aleatorias continuas que se usan para modelar un fenómeno con características que se evalúan en escalas continuas entonces:

$$p(y \mid x) = \frac{p(x \mid y)p(y)}{p(x)} = \frac{p(x \mid y)p(y)}{\int_Y p(x \mid y)p(y)dy} \propto p(x \mid y)p(y)$$

En donde de acuerdo con la terminología Bayesiana:

- $p(y)$ es la función de densidad *a priori* de *Y*.
- $p(y \mid x)$ es la función de densidad *a posteriori* de Y.
- El símbolo "$\propto$" se interpreta como "es proporcional a".

Nótese que se usa la notación $p(y)$ para la densidad a priori de *Y*, y $p(y|x)$ para la densidad condicional de *Y* dado $[X = x]$ (aunque, en general, son funciones diferentes), esta notación, aunque imprecisa, permite simplificar en mucho la exposición.

**El problema de inferencia paramétrica en el contexto Bayesiano**

El problema de inferencia paramétrica es un problema de decisión en el cual algún tomador de decisiones decide que una característica incierta $Y$ se puede modelar con la familia paramétrica de densidades $F = \{p(y|\theta): y \in \aleph, \theta \in \Theta \subset \Re\}$ y quiere o; decidir qué función de esta familia modela mejor el fenómeno (Estimación Puntual) o; decidir en que rango de valores es altamente probable que se encuentre $\theta$ tal que $p(y|\theta)$ sea la función de $F$ que modela mejor el fenómeno (Estimación por Regiones) o; comparar 2 posibles valores o regiones de valores para $\theta$ y decidir cual de las dos opciones contiene a $\theta$ tal que $p(y|\theta)$ sea la función de $F$ que modela mejor el fenómeno (Contraste de Hipótesis). Dicho de otra forma el tomador decisiones quiere hacer alguna inferencia sobre $\theta$. La única fuente de incertidumbre en el problema está en cuál es la función $p(y|\theta) \in F$ que mejor modela el fenómeno, es decir con qué valor del parámetro $\theta \in \Theta$ se puede modelar mejor la característica incierta, por lo que para resolver el problema el tomador de decisiones debe escoger una función de probabilidad $p(\theta)$ que refleje su incertidumbre con respecto al mejor valor de $\theta$. De acuerdo a la metodología Bayesiana el tomador de decisiones deberá escoger la función $p^*(y|\theta) \in F$, tal que $p^*(y|\theta)$ minimiza la esperanza, con respecto a $p(\theta)$, de la función de pérdida que el tomador de decisiones escogió. Si además se cuenta con una muestra aleatoria de observaciones del fenómeno $Y$ entonces se debe usar esta información a través del mecanismo del Teorema de Bayes para actualizar la medida de probabilidad de $\theta$. Si $x = \{y_1, y_2, ..., y_n\}$ es una muestra aleatoria de observaciones de $Y$ entonces:

$$p(\theta \mid x) = \frac{p(x \mid \theta)p(\theta)}{p(x)} = \frac{p(x \mid \theta)p(\theta)}{\int_\Theta p(x \mid \theta)p(\theta)d\theta} \propto p(x \mid \theta)p(\theta)$$

Cuando $x = \{y_1, y_2, ..., y_n\}$ es fijo y $\theta$ es desconocido $p(\theta \mid x)$ se conoce como la función de verosimilitud $L(\theta \mid x)$ que es una función de $\theta$. Entonces tenemos que:

$$p(\theta \mid x) = \frac{L(\theta \mid x)p(\theta)}{p(x)} = \frac{L(\theta \mid x)p(\theta)}{\int_\Theta L(\theta \mid x)p(\theta)d\theta} \propto L(\theta \mid x)p(\theta) = p(\theta)\prod_{i=1}^{n} p(y_i \mid \theta)$$

Notar que $L(\theta \mid x)p(\theta)$ es proporcional como función de $\theta$ a $p(\theta \mid x)$ por lo que para calcular $p(\theta \mid x)$ únicamente se necesita obtener $L(\theta \mid x)$ y $p(\theta)$.

## 1.2 Definición de la Problemática

### 1.2.1 Inicial de Referencia

Supóngase que se cuenta con un fenómeno que se puede modelar con algún elemento de la familia paramétrica de densidades $F = \{p(y \mid \theta): y \in \aleph, \theta \in \Theta \subset \Re\}$. Como ya se vio, para resolver el problema de inferencia paramétrica con la metodología Bayesiana, el tomador de decisiones debe proponer una función de densidad subjetiva a priori $p(\theta)$. Si se cuenta con una muestra aleatoria $x = \{y_1, y_2, ..., y_n\}$ el tomador de decisiones debe actualizar su medida de probabilidad de $\theta$ mediante el Teorema de Bayes:

$$p(\theta \mid x) = \frac{L(x \mid \theta)p(\theta)}{p(x)} = \frac{L(x \mid \theta)p(\theta)}{\int_\Theta L(x \mid \theta)p(\theta)} \propto L(x \mid \theta)p(\theta)$$

¿Pero qué hacer si el tomador de decisiones no cuenta con información alguna acerca del mejor valor para $\theta$? ¿Qué función de densidad a priori $p(\theta)$ refleja su incertidumbre acerca del mejor valor para $\theta$? ¿Habrá alguna función de densidad a priori $p(\theta)$ que represente ignorancia total acerca del mejor valor para $\theta$? Si además se cuenta con una muestra aleatoria del fenómeno que se quiere modelar, ¿habrá alguna función de densidad a priori $p(\theta)$ que, dado que no se cuenta con información alguna acerca de $\theta$ previa a la muestra, deje a los datos (a la muestra) tener la mayor influencia posible sobre la forma de $p(\theta|x)$, la función de densidad *a posteriori*?

Las funciones llamadas iniciales son estas funciones. Funciones que se usan en lugar de funciones de densidad iniciales subjetivas y que pretenden representar un estado inicial de ignorancia ó que los datos (la muestra) y no la subjetividad del tomador de decisiones, sean los que informen acerca de la probabilidad *a posteriori* de los posibles valores de $\theta$. Esta es una descripción de funciones iniciales de referencia muy vaga. Históricamente el problema de definir y proponer funciones a priori mínimo informativas se ha abordado de distintas maneras entre las cuales se encuentran el Principio de la razón insuficiente y las iniciales de referencia de Jeffreys [ver definición en Bernardo y Smith, 2000]. Basado en conceptos de la teoría de la información [Bernardo y Smith, 2000] propone una definición para las iniciales de referencia, y un método para obtenerlas. Sin embargo, resulta que en muchos casos es muy complicado obtenerlos analíticamente. Por esta razón [Berger et al., 2009] desarrollaron un método para estimarlas por medio de simulación. Este método es el punto de partida de esta investigación. En el capítulo 2 se explorará el método desarrollando 3 ejemplos de implementación. En el capítulo 3 se desarrollarán medidas para estimar el error de las estimaciones que se producen por este método. En el capítulo 4 se aplicará una técnica de reducción de varianza a los estimadores de este método en un intento de disminuir su error de estimación.

En el contexto de la Ingeniería Industrial el problema de inferencia paramétrica se presenta en áreas como Control de Calidad, Planeación de Proyectos e Investigación de Operaciones. En el Control de Calidad generalmente interesa verificar que cierta característica de un producto o servicio, el peso de una pelota de golf o la acidez de algún producto químico por ejemplo, se mantenga dentro de ciertos rangos, y para hacerlo se propone usar una familia paramétrica de densidades para modelar la incertidumbre que se tiene con respecto a la característica de interés, se toman muestras del producto o servicio, midiendo la característica de interés y se hace inferencia sobre los parámetros que especifican la función que se usará y que permitirá concluir sobre la efectividad del proceso relevante en producir productos o servicios de cierta calidad. Uno de los problemas fundamentales de la Planeación de Proyectos es establecer cuales de las actividades que conforman un proyecto son críticas para la duración del proyecto (ruta crítica). A menudo se usa las funciones de densidad triangular o beta para modelar la incertidumbre que se tiene con respecto a la duración de una o varias de las actividades del proyecto. Sí se cuenta con una muestra de la duración de cierta actividad en el pasado se puede utilizar dicha muestra para hacer inferencia sobre los parámetros que definen la función especifica de la familia triangular o beta más adecuada para modelar la duración de la actividad. Dentro de la Investigación de Operaciones, la teoría de colas estudia el fenómeno de de las líneas de espera. En esta teoría, a menudo se utilizan las funciones de densidad exponencial y poisson para modelar el tiempo entre llegadas y el número de llegadas por unidad de tiempo. Se pueden utilizar datos históricos de las llegadas de clientes a una línea de espera para hacer inferencia sobre el parámetro $\lambda$ que relaciona a la distribución exponencial y a la poisson y que representa la tasa media de llegadas.

En todos estos problemas es común que el tomador de decisiones no tenga conocimiento alguno, aparte de la muestra con la que cuenta, sobre el mejor valor para el parámetro que especificará la función de densidad con la que describirá su incertidumbre con respecto al evento de interés. Las funciones iniciales de referencia representan una solución al problema de escoger la mejor función dentro de una familia paramétrica de densidades, para modelar la incertidumbre con respecto a cierto evento, basándose únicamente en la información que provee una muestra de repeticiones del evento relevante.

**1.2.2 Notación**

- $X_j^{(k)} = (Y_{1j}, \ldots, Y_{kj})$ es un vector de variables aleatorias independiente e idénticamente distribuidas con función de densidad $p(y|\theta)$

- $p(x|\theta) = \prod_{i=1}^{k} p(y|\theta)$ es la función de densidad conjunta del vector $X_j^{(k)}$

- $x_j^{(k)} = (y_{1j}, \ldots, y_{kj})$ es una observación del vector aleatorio $X_j^{(k)}$.

- Si se evalúa $p(x|\theta) = \prod_{i=1}^{k} p(y|\theta)$ en $x_j^{(k)} = (y_{1j}, \ldots, y_{kj})$ la función de densidad conjunta es la verosimilitud del vector de observaciones $x_j^{(k)}$, una función en $\theta$.

- Un modelo $M = \{p(x|\theta) : x \in \aleph^k, \theta \in \Theta \subset \Re\}$ se define como la situación en la que se estudia un fenómeno que se ha decidido modelar con una función de la familia paramétrica $F = \{p(y|\theta) : y \in \aleph, \theta \in \Theta \subset \Re\}$ y del cual se cuenta con un vector de observaciones $x_j^{(k)} = (y_{1j}, \ldots, y_{kj})$ con función de densidad conjunta $p(x|\theta)$.

- $f(\theta)$, es proporcional a la inicial de referencia propuesta en Berger et al. [2009] correspondiente al modelo $M = \{p(x|\theta): x \in \aleph^k, \theta \in \Theta \subset \Re\}$.

- $t_k$ es una estadística suficiente para $\theta$, de $x_j^{(k)}$

- $f_k(\theta) = \exp\left\{\int p(t_k|\theta)\left[\log[\pi^*(\theta|t_k)]\right]dt_k\right\}$ donde $\pi^*(\theta|t_k) = \dfrac{p(t_k|\theta)\pi^*(\theta)}{\int_\Theta p(t_k|\theta)\pi^*(\theta)d\theta}$ se asume propia y consistente (cuando $k \to \infty$).

- $\pi^*(\theta)$ es una función inicial continua positiva arbitraria.

- Si escogemos $\pi^*(\theta) = 1$ y $t_k = x_j^{(k)}$ (la estadística suficiente es toda la muestra) entonces $f_k(\theta) = \exp\left\{E_{p(x^{(k)}|\theta)}\left[\log\left[\dfrac{p(x^{(k)}|\theta)}{\int_\Theta p(x^{(k)}|\theta)d\theta}\right]\right]\right\}$

- $\hat{f}_k(\theta) = \dfrac{1}{m}\sum_{j=1}^{m}\log\left[\dfrac{p(x_j^{(k)}|\theta)}{\int_\Theta p(x_j^{(k)}|\theta)d\theta}\right]$ y si definimos $c_j = \int_\Theta p(x_j^{(k)}|\theta)d\theta$ y

  $r_j(\theta) = \log(p(x_j^{(k)}|\theta)/c_j)$ entonces $\hat{f}_k(\theta) = \exp\left[\sum_{j=1}^{k} r_j(\theta)/k\right]$

- $\dfrac{\hat{f}(\theta)}{\hat{b}}$ y $\dfrac{\hat{f}_k(\theta)}{\hat{a}}$ son dos estimadores de $f(\theta)$

**1.3 Preguntas de Investigación**

Las preguntas de investigación se presentan a continuación en orden de importancia:

- ¿Es adecuado el método propuesto por Berger et al. [2009] para calcular la inicial de referencia?
- ¿Cómo se puede estimar el error de la estimación (por simulación) de la inicial de referencia?
- ¿Es posible disminuir el error de los estimadores utilizando técnicas de reducción de varianza?

**1.4 Objetivos**

Los objetivos de esta investigación se presentan a continuación en orden de importancia:

1. Distinguir cuál de los dos estimadores de la inicial de referencia, estudiados en este trabajo, tiene mejor desempeño con respecto a medidas como cobertura empírica y error de estimación.
2. Evaluar los distintos estimadores estudiados en este trabajo en cuanto a eficiencia: ¿Cuántas simulaciones requiere cada uno para obtener un error aceptable?
3. Formular una metodología para evaluar la precisión de los estimadores y así poder especificar el tamaño y número de muestras requerido para que el método propuesto en Berger et al. [2009] resulte en estimadores de la inicial de referencia con un cierto nivel de precisión preestablecido.
4. Investigar si usando números aleatorios comunes se puede reducir la varianza de los estimadores de la inicial de referencia. Distinguir en qué casos sí se consiguen buenos resultados con esta técnica y en qué casos no.

# 2. CÁLCULO DE LA DISTRIBUCIÓN INICIAL DE REFERENCIA PARA $\theta$, POR SIMULACIÓN

## 2.1 Teorema Principal

Antes de presentar el teorema principal del texto de Berger et al. [2009] se presentan algunas definiciones que se usaran en dicho teorema y que se toman del mismo texto.

**Divergencia logarítmica (Divergencia Kullback-Leibler).** La divergencia logarítmica de una función de densidad $\tilde{p}(\theta)$ de la variable aleatoria $\theta \in \Theta$, de su verdadera función de densidad $p(\theta)$, denotada $\kappa\{\tilde{p} \mid p\}$, es

$$\kappa\{\tilde{p} \mid p\} = \int_\Theta p(\theta) \log \frac{p(\theta)}{\tilde{p}(\theta)} d\theta$$

**Convergencia logarítmica.** Una secuencia de funciones de densidad $\{p_i(\theta)\}_{i=1}^{\infty}$ converge de forma logarítmica a una función de densidad $p(\theta)$ si, y solo si, $\lim_{i \to \infty} \kappa\{p \mid p_i\} = 0$

**Información Esperada**. La información esperada de una observación del modelo $M = \{p(x \mid \theta): x \in \aleph^k, \theta \in \Theta \subset \Re\}$, cuando la inicial para $\theta$ es $q(\theta)$, es:

$$I\{q \mid M\} = \int_{\aleph^k x \Theta} p(x \mid \theta) q(\theta) \log \frac{p(\theta \mid x)}{q(\theta)} dx d\theta$$

**Funciones iniciales estándar**. Sea $P_s$ la clase de funciones iniciales estrictamente positivas y continuas en $\Theta$ que tienen funciones *a posteriori* propias de tal manera que, si $p \in P_s$ entonces,

$$\forall \theta \in \Theta, p(\theta) > 0; \quad \forall x \in \aleph^k, \int_\Theta p(x \mid \theta) p(\theta) d\theta < \infty .$$

**Modelo estándar**. Sea $M = \{p(x|\theta): x \in \aleph^k, \theta \in \Theta \subset \Re\}$ un modelo con parámetro $\theta$ en escala continua y suponer que se cuentan con k repeticiones de este modelo, es decir con $z^{(k)} = (x_1,...,x_k)$. Se dice que el modelo *M* es estándar si, para cualquier función inicial $p(\theta) \in P_s$ y cualquier conjunto compacto $\Theta_0$;

$I\{p_0 | M\} < \infty$, donde $p_0(\theta)$ es la función inicial propia que se obtiene cuando se restringe $p(\theta)$ a $\Theta_0$

**Secuencia compacta aproximante.** Considerar un modelo paramétrico $M = \{p(x|\theta): x \in \aleph^k, \theta \in \Theta \subset \Re\}$ y una función estrictamente positiva y continua $\pi(\theta), \theta \in \Theta$ tal que, $\forall x \in \aleph^k, \int_\Theta p(x|\theta)p(\theta)d\theta < \infty$. Una secuencia compacta aproximante de espacios parametrales es una secuencia creciente de subconjuntos compactos de $\Theta, \{\Theta_i\}_{i=1}^\infty$, que convergen a $\Theta$. La secuencia correspondiente de funciones *a posteriori* con soporte en $\Theta_i$, definidos como $\{\pi_i(\theta|x)\}_{i=1}^\infty$, con $\pi_i(\theta|x) \propto p(x|\theta)\pi_i(\theta)$, $\pi_i(\theta) = c_i^{-1}\pi(\theta)$ si $\theta \in \Theta_i$ y 0 en otro caso, y $c_i = \int_{\Theta_i} \pi(\theta)d\theta$, se llama la secuencia aproximante de funciones *a posteriori* a la función *a posteriori* formal $\pi(\theta|x)$.

**Convergencia logarítmica esperada de funciones *a posteriori*.** Considerar un modelo paramétrico $M = \{p(x|\theta): x \in \aleph^k, \theta \in \Theta \subset \Re\}$, una función estrictamente positiva y continua $\pi(\theta), \theta \in \Theta$ y una secuencia compacta aproximante $\{\Theta_i\}$ de espacios parametrales. La secuencia correspondiente de funciones *a posteriori* $\{\pi_i(\theta|x)\}_{i=1}^\infty$ se dice que se espera que converga de forma logarítmica a la función *a posteriori* formal $\pi(\theta|x)$ si

$$\lim_{i \to \infty} \int_{\aleph^k} \kappa\{\pi(\theta \mid x) \mid \pi_i(\theta \mid x)\} p_i(x) dx = 0 \text{ donde } p_i(x) dx = \int_{\Theta_i} p(x \mid \theta) \pi_i(\theta) d\theta$$

**Función inicial permisible**. Una función estrictamente positiva $\pi(\theta)$ es una inicial permisible para un modelo $M = \{p(x \mid \theta): x \in \aleph^k, \theta \in \Theta \subset \Re\}$ si:

1. $\forall x \in \aleph^k, \int_{\Theta} p(x \mid \theta) p(\theta) d\theta < \infty$

2. Para alguna secuencia compacta aproximante, la secuencia correspondiente de funciones *a posteriori* se espera que converja de forma logarítmica a $\pi(\theta \mid x) \propto p(x \mid \theta) \pi(\theta)$

**Consistencia de una función *a posteriori***. Una función *a posteriori* $\pi^*$ es consistente si para toda $\theta \in \Theta$ y $\varepsilon > 0$,

$$P(\mid \tau - \theta \mid \leq \varepsilon \mid t_k) \equiv \int_{\{\tau: \mid \tau - \theta \mid \leq \varepsilon\}} \pi^*(\tau \mid t_k) d\tau \xrightarrow{P} 1 \text{ cuando } k \to \infty$$

A continuación se presenta el resultado principal presentado en el texto de Berger et al. [2009] referente a la definición y procedimiento para construir la inicial de referencia $f(\theta)$ correspondiente a cierto modelo $M = \{p(x|\theta): x \in \aleph^k, \theta \in \Theta \subset \Re\}$.

**Teorema: Forma explícita de la inicial de referencia**. Asumir un modelo estándar $M \equiv \{p(x|\theta): x \in \aleph^k, \theta \in \Theta \subset \Re\}$, y la clase estándar $P_s$ de iniciales candidatas. Sea $\pi^*(\theta)$ una función continua y estrictamente positiva tal que la *a posteriori* formal

$$\pi^*(\theta|t_k) = \frac{p(t_k|\theta)\pi^*(\theta)}{\int_\Theta p(t_k|\theta)\pi^*(\theta)d\theta}$$

es propia y consistente, y definir para cualquier punto interior $\theta_0$ de $\Theta$,

$$f_k(\theta) = \exp\left\{\int p(t_k|\theta)\log[\pi^*(\theta|t_k)]dt_k\right\}, \text{ y}$$

$$f(\theta) = \lim_{k\to\infty} \frac{f_k(\theta)}{f_k(\theta_0)}$$

Si (i) $f_k(\theta)$ es continua para toda $k$ y $t_k$ y, para cualquier $\theta$ fija y $k$ suficientemente grande, $\left\{\dfrac{f_k(\theta)}{f_k(\theta_0)}\right\}$ es o monótona en k o está acotada por arriba por alguna función $h(\theta)$ que es integrable en cualquier conjunto compacto; y (ii) $f(\theta)$ es una función inicial permisible, entonces $\pi(\theta|M,P_s) = f(\theta)$ es una inicial de referencia para el modelo M y la clase de iniciales $P_s$.

El teorema muestra la forma explícita de la inicial de referencia [Berger et al., 2009, p15] con base en el cual se formuló el método de estimación de la inicial de referencia por simulación. Lo que este teorema enuncia es que si $\theta_0$ es un punto interior del espacio de parámetros $\Theta$ entonces para cualquier $\theta \in \Theta$, la inicial de referencia $f(\theta)$ cumple: $f(\theta) = \lim_{k \to \infty} \frac{f_k(\theta)}{f_k(\theta_0)}$. Si escogemos a la inicial continua positiva arbitraria $\pi^*(\theta) = 1$ y usamos como estadística suficiente a la muestra entera ($t_k = x_j^{(k)}$), entonces tenemos que

$$f_k(\theta) = \exp\left\{ E_{p(x^{(k)}|\theta)}\left[ \log\left[ \frac{p(x^{(k)}|\theta)}{\int_{\Re^k} p(x^{(k)}|\theta) d\theta} \right] \right] \right\} .$$

Un resultado inmediato del teorema anterior es la posibilidad de estimar por medio de simulación la inicial de referencia, lo que resulta útil para modelos $M \equiv \{p(x|\theta): x \in \aleph^k, \theta \in \Theta \subset \Re\}$ para los cuales el procedimiento analítico para construir la inicial de referencia $f(\theta)$, presentado en el teorema anterior, es complicado. A continuación se muestra el procedimiento incluido en Berger et al. [2009] para estimar, por medio de simulación, la inicial de referencia correspondiente a un modelo $M \equiv \{p(x|\theta): x \in \aleph^k, \theta \in \Theta \subset \Re\}$.

**Algoritmo para estimar la inicial de referencia**

1. Valores iniciales:

    Escoger un valor moderado para k;

    Escoger una función positiva arbitraria $\pi^*(\theta)$, por ejemplo $\pi^*(\theta)=1$;

    Escoger el número m de muestras a simular.

2. Para algún valor $\theta$, repetir lo siguiente para j=1,…,m:

    Simular una muestra aleatoria $x_j^{(k)} = (y_{1j},\ldots, y_{kj})$ de tamaño k de $p(x|\theta)$;

    Calcular con métodos numéricos la integral $c_j = \int \prod_{i=1}^{k} p(y_{ij}|\theta)\pi^*(\theta)d\theta$;

    Evaluar $r_j(\theta) = \log\left[\prod_{i=1}^{k} \frac{p(y_{ij}|\theta)\pi^*(\theta)}{c_j}\right]$

3. Calcular $\hat{f}_k(\theta) = \exp\left[m^{-1}\sum_{j=1}^{m} r_j(\theta)\right]$ y guardar la pareja $\{\theta, \hat{f}_k(\theta)\}$

4. Repetir rutinas (2) y (3) para todos los valores de $\theta$ para los cuales la pareja $\{\theta, \hat{f}_k(\theta)\}$ es requerida.

El método de estimación de la inicial de referencia de Berger et al. [2009, p17] consiste en estimar la expresión $f(\theta) = \lim_{k\to\infty} \frac{f_k(\theta)}{f_k(\theta_0)}$ con $\hat{f}_k(\theta)$ para k suficientemente grande. La expresión $f_k(\theta) = \exp\left\{E_{p(x^{(k)}|\theta)}\left[\log\left[\frac{p(x^{(k)}|\theta)}{\int_\Theta p(x^{(k)}|\theta)d\theta}\right]\right]\right\}$ a su vez se estima simulando $m$ muestras aleatorias de tamaño $k$ con función de densidad conjunta $p(x_j^{(k)}|\theta) = \prod_{i=1}^{k} p(y_{ij}|\theta)$ y estimando la expresión $E_{p(x^{(k)}|\theta)}\left[\log\left[\dfrac{p(x^{(k)}|\theta)}{\int_\Theta p(x^{(k)}|\theta)d\theta}\right]\right]$ con $\dfrac{1}{m}\sum_{j=1}^{m}\log\left[\dfrac{p(x_j^{(k)}|\theta)}{\int_\Theta p(x_j^{(k)}|\theta)d\theta}\right]$ para

$k$ y $m$ suficientemente grandes. Con el fin de simplificar el análisis se considera que $m=k$, de manera que se simulan $k$ muestras de tamaño $k$. Si definimos $c_j = \int_\Theta p(x_j^{(k)}|\theta)d\theta$ y

$r_j(\theta) = \log(p(x_j^{(k)}|\theta)/c_j)$ entonces el estimador resulta ser $\hat{f}_k(\theta) = \exp\left[\sum_{j=1}^{k} r_j(\theta)/k\right]$.

Asumiendo que $\lim_{k\to\infty} f_k(\theta)$ converge para toda $\theta$ y $\lim_{k\to\infty} f_k(\theta_0)$ converge, $f_k(\theta_0)$ es una constante y $\hat{f}_k(\theta_0)$ aproxima está constante. Por lo tanto para estimar una inicial de referencia podemos:

1. Calcular $\hat{f}_k(\theta)$

2. Calcular $\hat{f}(\theta) = \dfrac{\hat{f}_k(\theta)}{\hat{f}_k(\theta_0)}$

Si $f(\theta)$ es una función inicial de referencia, entonces $cf(\theta)$, donde $c$ es una constante positiva, también es una función inicial de referencia. En realidad $\hat{f}_k(\theta)$ es un estimador de la inicial de referencia $g(\theta) = af(\theta)$ y $\hat{f}(\theta)$ es un estimador de la inicial de referencia $h(\theta) = bf(\theta)$ donde $a$ y $b$ son constantes positivas. En general, no importa que función inicial de referencia de la familia $P = \left\{cf(\theta): c>0, \theta \in \Theta, f(\theta) \propto \lim_{k\to\infty}\dfrac{f_k(\theta)}{f_k(\theta_0)}\right\}$ se escoja puesto que una vez multiplicadas por la función de verosimilitud y normalizadas todas resultaran en la misma función de densidad *a posteriori*.

$$p(\theta \mid x) = \frac{L(x \mid \theta) c f(\theta)}{\int_{\Theta} L(x \mid \theta) c f(\theta)} \propto L(x \mid \theta) f(\theta)$$

Sin embargo, en los ejemplos de implementación, para poder comparar la precisión de ambos estimadores, propondremos una función inicial de referencia $f(\theta)$ que se ha comprobado que es proporcional a $\lim_{k \to \infty} \frac{f_k(\theta)}{f_k(\theta_0)}$ y estimaremos las constantes $a$ y $b$ de tal manera que $\frac{\hat{f}_k(\theta)}{\hat{a}}$ y $\frac{\hat{f}(\theta)}{\hat{b}}$ sean estimadores de $f(\theta)$.

A primera vista parece redundante estimar $\hat{f}(\theta)$ pues requiere la estimación adicional de $\hat{f}_k(\theta_0)$, que estima a una constante. Es importante recordar que lo que se busca con esté método en general, es estimar $f(\theta)$ para distintos valores de $\theta$ de forma que se pueda aproximar la forma algebraica que tiene $f(\theta)$ y usar esta inicial, en conjunto con la función de verosimilitud para calcular la función *a posteriori* de probabilidad. En la estimación de $f(\theta)$ con $\frac{\hat{f}(\theta)}{\hat{b}} = \frac{\hat{f}_k(\theta)}{\hat{f}_k(\theta_0)} \frac{1}{\hat{b}}$ para cada valor de $\theta$ para el cual se busca $f(\theta)$ estaremos usando una muestra aleatoria para estimar $\hat{f}_k(\theta)$ y otra para estimar $\hat{f}_k(\theta_0)$ por lo que aunque es cierto que $\frac{f_k(\theta)}{f_k(\theta_0)} \propto f_k(\theta)$ y por lo tanto $\hat{f}(\theta)$ y $\hat{f}_k(\theta)$ estiman a iniciales de referencia proporcionales entre sí, no existe una constante c tal que para toda $\theta$ para la cual se calcule $\hat{f}_k(\theta)$ y $\hat{f}(\theta)$ sea cierto que $\hat{f}(\theta) = c\hat{f}_k(\theta)$. A continuación se ilustra, con el ejemplo de datos con distribución uniforme entre 0 y $\theta$, porque tiene cierto sentido estimar tanto $\hat{f}_k(\theta)$ como $\hat{f}(\theta)$, y se aprovecha para introducir y ejemplificar las medidas que se utilizaran para evaluar el desempeño de ambos estimadores.

**Distribución Uniforme** $(0, \theta)$

El modelo está definido por:

$$p(y|\theta) = \frac{1}{\theta}, 0 < y < \theta, 0 < \theta < \infty$$

$$p(x_j^{(k)}|\theta) = \prod_{i=1}^{k} p(y_{ij}|\theta) = \frac{1}{\theta^k} \prod_{i=1}^{k} I_{(0,\theta)}(y_{ij}) = \frac{1}{\theta^k} I_{(t_{(k)j},\infty)}(\theta) \text{ con } t_{(k)j} = \max_{i=1..k}\{y_{ij}\}$$

$$c_j = \int_\Theta p(x_j^{(k)}|\theta) d\theta = \int_{t_{(k)j}}^{\infty} \theta^{-k} d\theta = \frac{t_{(k)j}^{1-k}}{(k-1)}$$

$$r_j(\theta) = \log(p(x_j^{(k)}|\theta)/c_j) = \log\left(\frac{k-1}{\theta^k t_{(k)j}^{1-k}}\right) = \log(k-1) - k\log(\theta) + (k-1)\log(t_{(k)j})$$

Se ejemplificara el método estimando $\hat{f}_k(\theta)$ y $\hat{f}_k(\theta_0)$ para $\theta = 5$, con k=5 y $\theta_0 = 1$:

**Tabla 2.1: Cálculo de $\hat{f}_k(\theta)$ para datos distribuidos $Unif(0, \theta)$**

Simulación de 5 muestras de tamaño 5 para estimación de fk(θ)

| No. de muestra (j) | \multicolumn{5}{c}{No. de observación (i)} | | | | | Máximo | cj | rj(θ) |
|---|---|---|---|---|---|---|---|---|
| | 1 | 2 | 3 | 4 | 5 | | | |
| 1 | 2.643036 | 2.525562 | 0.960058 | 4.832099 | 4.272201 | 4.832099 | 0.000459 | -0.359772 |
| 2 | 2.174483 | 2.483448 | 1.491607 | 4.914156 | 0.174570 | 4.914156 | 0.000429 | -0.292415 |
| 3 | 1.941754 | 1.177051 | 1.256304 | 4.244871 | 2.803651 | 4.244871 | 0.000770 | -0.878049 |
| 4 | 0.451879 | 2.500297 | 4.722214 | 4.968808 | 2.783378 | 4.968808 | 0.000410 | -0.248175 |
| 5 | 3.850290 | 1.067490 | 3.773885 | 1.241600 | 1.111622 | 3.850290 | 0.001138 | -1.268302 |

**Tabla 2.2: Cálculo de $\hat{f}_k(\theta_0)$ para datos distribuidos $Unif(0, \theta)$**

Simulación de 5 muestras de tamaño 5 para estimación de fk(θ0)

| No. de muestra (j) | \multicolumn{5}{c}{No. de observación (i)} | | | | | Máximo | cj | rj(θ0) |
|---|---|---|---|---|---|---|---|---|
| | 1 | 2 | 3 | 4 | 5 | | | |
| 1 | 0.302285 | 0.423168 | 0.138452 | 0.616580 | 0.575441 | 0.616580 | 1.729750 | -0.547977 |
| 2 | 0.307996 | 0.862337 | 0.886713 | 0.442853 | 0.799809 | 0.886713 | 0.404397 | 0.905358 |
| 3 | 0.011259 | 0.539374 | 0.939005 | 0.709738 | 0.193020 | 0.939005 | 0.321565 | 1.134554 |
| 4 | 0.947076 | 0.498836 | 0.251442 | 0.152291 | 0.045622 | 0.947076 | 0.310743 | 1.168789 |
| 5 | 0.364147 | 0.260889 | 0.536815 | 0.514442 | 0.604568 | 0.604568 | 1.871367 | -0.626669 |

Entonces utilizando $\hat{f}_k(\theta) = \exp\left[\sum_{j=1}^{k} r_j(\theta)/k\right]$:

1. $\hat{f}_k(\theta) = 0.5437$

2. $\hat{f}(\theta) = \dfrac{\hat{f}_k(\theta)}{\hat{f}_k(\theta_0)} = 0.5437/1.5020 = 0.3619$

Repetimos el proceso, ahora calculando ambos estimadores para $\theta = 2, 5, 8, 11, 14$ y $17$ con k=50 y $\theta_0 = 1$. Se sabe que la inicial de referencia para datos con función de densidad $p(y|\theta) \sim Unif(0,\theta)$ es proporcional a $\theta^{-1}$ [Bernardo, 2005].

**Tabla 2.3: Cálculo de $\hat{f}_k(\theta)$ y $\hat{f}(\theta)$ para datos distribuidos $Unif(0,\theta)$**

| θ | f(θ) | f(θ)gorro | f$_k$(θ)gorro | f(θ)gorro/f$_k$(θ)gorro |
|---|------|-----------|---------------|--------------------------|
| 2 | 0.500 | 0.563 | 9.941 | 0.057 |
| 5 | 0.200 | 0.198 | 4.047 | 0.049 |
| 8 | 0.125 | 0.098 | 1.907 | 0.051 |
| 11 | 0.091 | 0.062 | 1.397 | 0.044 |
| 14 | 0.071 | 0.082 | 1.455 | 0.057 |
| 17 | 0.059 | 0.055 | 1.095 | 0.051 |

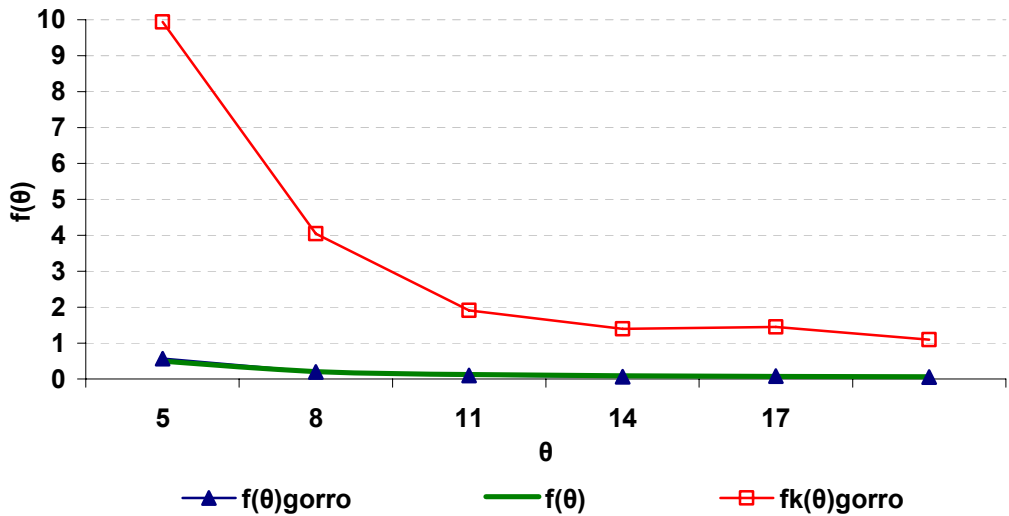

**Figura 2.1: Estimación de $f_k(\theta)$ y $f(\theta)$ para datos distribuidos $Unif(0,\theta)$**

Se puede ver que $\hat{f}_k(\theta)$ y $\hat{f}(\theta)$ están estimando funciones iniciales de referencia que difieren por alguna constante de proporcionalidad $c$. Para poder comparar el error de ambos estimadores habrá que estimar las constantes $a$ y $b$ de tal manera que $\dfrac{\hat{f}_k(\theta)}{\hat{a}}$ y $\dfrac{\hat{f}(\theta)}{\hat{b}}$ sean estimadores de $f(\theta)$. Además notar en la tabla 2.3 que $\hat{f}_k(\theta)$ y $\hat{f}(\theta)$ no difieren por una misma constante por lo que, en un sentido es razonable comparar los estimadores para ver cual de los dos se desempeña mejor. En otro sentido parece obvio que, dado que el estimador $\hat{f}(\theta)$ se ve afectado tanto por el error de estimación del numerador como del denominador, el estimador $\hat{f}_k(\theta)$ será más preciso. Sin embargo se mostrará que el estimador $\hat{f}(\theta)$ tiene algunas ventajas sobre el estimador $\hat{f}_k(\theta)$, como es la posibilidad de usar números aleatorios comunes para reducir la varianza del estimador.

Para poder comparar la precisión de ambos estimadores estimamos la constante $a$ y $b$ de tal forma que se minimice el error absoluto relativo promedio (EARP) entre $\dfrac{\hat{f}_k(\theta)}{a}$ y $f(\theta)$ y entre $\dfrac{\hat{f}(\theta)}{b}$ y $f(\theta)$. Si se calcula $\dfrac{\hat{f}(\theta)}{b}$ y $\dfrac{\hat{f}_k(\theta)}{a}$ para R distintos valores de $\theta$ el EARP de ambos estimadores se define como sigue:

$$EARP\left(\dfrac{\hat{f}(\theta)}{b}\right) = \sum_{l=1}^{R} \dfrac{\left|bf(\theta_l) - \hat{f}(\theta_l)\right|}{bf(\theta_l)R} \text{ y } EARP\left(\dfrac{\hat{f}_k(\theta)}{a}\right) = \sum_{l=1}^{R} \dfrac{\left|af(\theta_l) - \hat{f}_k(\theta_l)\right|}{af(\theta_l)R}$$

En la práctica buscamos estimar $f(\theta)$ para varios valores de $\theta$ y poder así estimar, por medio de técnicas de interpolación estándar, la forma continua de la inicial y usar está en el proceso de actualización Bayesiana cuyo fin es obtener la función de densidad *a posteriori*. Por está razón el EARP evalúa el desempeño de los estimadores en varios valores de $\theta$ simultáneamente.

El EARP no es una función continúa en $a$ por lo que no se puede encontrar el punto mínimo calculando la derivada con respecto a $a$, igualando a cero y resolviendo la ecuación resultante. Tomando el ejemplo de $EARP\left(\dfrac{\hat{f}_k(\theta)}{a}\right)$ si re-ordenamos y re-indexamos las R observaciones de $\hat{f}(\theta)$ y $\hat{f}_k(\theta)$ de tal forma que:

i) $a\hat{f}_k(\theta_i) > f(\theta_i)$ para $i = 0,...,s$

ii) $a\hat{f}_k(\theta_i) < f(\theta_i)$ para $i = s+1,...,R$,

iii) $\dfrac{\hat{f}_k(\theta_i)}{f(\theta_i)} < \dfrac{\hat{f}_k(\theta_{i+1})}{f(\theta_{i+1})}$ para $i = 1,...,R-1$, entonces tenemos que

$$EARP\left(\dfrac{\hat{f}_k(\theta)}{a}\right) = \sum_{i=1}^{R}\left[\dfrac{1}{a}\dfrac{\hat{f}_k(\theta_i)}{f(\theta_i)R} - \dfrac{1}{R}\right] = \dfrac{1}{aR}\sum_{i=1}^{R}\left[\dfrac{\hat{f}_k(\theta_i)}{f(\theta_i)}\right] - 1 \text{ si s = 0}$$

$$EARP\left(\dfrac{\hat{f}_k(\theta)}{a}\right) = \sum_{i=1}^{s}\left[\dfrac{1}{R} - \dfrac{1}{a}\dfrac{\hat{f}_k(\theta_i)}{f(\theta_i)R}\right] + \sum_{i=s+1}^{R}\left[\dfrac{1}{a}\dfrac{\hat{f}_k(\theta_i)}{f(\theta_i)R} - \dfrac{1}{R}\right]$$

$$= \dfrac{2s}{R} - 1 + \dfrac{1}{aR}\left[\sum_{i=s+1}^{R}\dfrac{\hat{f}_k(\theta_i)}{f(\theta_i)} - \sum_{i=1}^{s}\dfrac{\hat{f}_k(\theta_i)}{f(\theta_i)}\right] \text{.si } s \in \{1,...,R-1\}$$

y $EARP\left(\dfrac{\hat{f}_k(\theta)}{a}\right) = \sum_{i=1}^{R}\left[\dfrac{1}{R} - \dfrac{1}{a}\dfrac{\hat{f}_k(\theta_i)}{f(\theta_i)R}\right] = 1 - \dfrac{1}{aR}\sum_{i=1}^{R}\left[\dfrac{\hat{f}_k(\theta_i)}{f(\theta_i)}\right]$ si $s = R$

Por (i), (ii) y (iii) $0 < \dfrac{\hat{f}_k(\theta_0)}{f(\theta_0)} < \dfrac{\hat{f}_k(\theta_1)}{f(\theta_1)} < ... < \dfrac{\hat{f}_k(\theta_s)}{f(\theta_s)} < a < \dfrac{\hat{f}_k(\theta_{s+1})}{f(\theta_{s+1})} < ... < \dfrac{\hat{f}_k(\theta_R)}{f(\theta_R)}$,

$s \in \{0,...,R\}$, por lo que para $s$ fija el EARP se minimiza escogiendo $\hat{a} = \dfrac{\hat{f}_k(\theta_1)}{f(\theta_1)}$ si $s = 0$,

$\hat{a} = \dfrac{\hat{f}_k(\theta_{s+1})}{f(\theta_{s+1})}$ si $s \in \{1,...,R-1\}$ y escogiendo $\hat{a} = \dfrac{\hat{f}_k(\theta_R)}{f(\theta_R)}$ si $s = R$ de tal forma que para encontrar $\hat{a}$ tal que el EARP sea mínimo únicamente falta seleccionar $\hat{s} \in \{0,...,R\}$ tal que

$$EARP\left(\frac{\hat{f}_k(\theta)}{a}\right) = \left(\frac{f(\theta_1)}{\hat{f}_k(\theta_1)R}\sum_{i=1}^{R}\left[\frac{\hat{f}_k(\theta_i)}{f(\theta_i)}\right] - 1\right)I_{\{0\}}(s) +$$

$$\left(\frac{2s}{R} - 1 + \frac{f(\theta_{s+1})}{\hat{f}_k(\theta_{s+1})R}\left[\sum_{i=s+1}^{R}\frac{\hat{f}_k(\theta_i)}{f(\theta_i)} - \sum_{i=1}^{s}\frac{\hat{f}_k(\theta_i)}{f(\theta_i)}\right]\right)I_{\{1,\ldots,R-1\}}(s) + \left(1 - \frac{f_R(\theta_R)}{\hat{f}_R(\theta_R)R}\sum_{i=1}^{R}\left[\frac{\hat{f}_k(\theta_i)}{f(\theta_i)}\right]\right)I_{\{R\}}(s)$$

sea mínimo. Esto se puede realizar calculando el EARP para $s = 0,\ldots,R$ y seleccionando la $s$ que minimice el resultado.

Siguiendo con el ejemplo a partir de datos con distribución uniforme entre $0$ y $\theta$, se calculan ambos estimadores para $\theta = 2, 5, 8, 11, 14$ y $17$ con k=50 y $\theta_0 = 1$ y se calculan las constantes $a$ y $b$.

**Tabla 2.4: Estimación de la constante de proporcionalidad de *a* para datos distribuidos** $Unif(0,\theta)$

| | | | Cálculo de a | | | | |
|---|---|---|---|---|---|---|---|
| $\theta$ | $f(\theta)$ | $f_k(\theta)$gorro | s | fk($\theta$)gorro/ f($\theta$) | fk($\theta$)gorro/ f($\theta$) ordenado | EARP (fk($\theta$)gorro) | $f_k(\theta)$gorro/a |
| | | | 0 | | | 0.198 | |
| 2 | 0.500 | 9.941 | 1 | 19.883 | 15.260 | 0.192 | 0.500 |
| 5 | 0.200 | 4.047 | 2 | 20.233 | 15.366 | 0.101 | 0.204 |
| 8 | 0.125 | 1.907 | 3 | 15.260 | 18.622 | 0.094 | 0.096 |
| 11 | 0.091 | 1.397 | 4 | 15.366 | 19.883 | 0.098 | 0.070 |
| 14 | 0.071 | 1.455 | 5 | 20.365 | 20.233 | 0.102 | 0.073 |
| 17 | 0.059 | 1.095 | 6 | 18.622 | 20.365 | 0.102 | 0.055 |

**Tabla 2.5: Estimación de la constante de proporcionalidad de *b* para datos distribuidos** $Unif(0,\theta)$

| | | | | Cálculo de b | | | |
|---|---|---|---|---|---|---|---|
| θ | f(θ) | f(θ)gorro | s | f(θ)gorro/ f(θ) | f(θ)gorro/ f(θ) ordenado | EARP (f(θ)gorro) | f(θ)gorro/b |
| | | | 0 | | | 0.387 | |
| 2 | 0.500 | 0.563 | 1 | 1.126 | 0.683 | 0.251 | 0.568 |
| 5 | 0.200 | 0.198 | 2 | 0.991 | 0.784 | 0.152 | 0.200 |
| 8 | 0.125 | 0.098 | 3 | 0.784 | 0.943 | 0.145 | 0.099 |
| 11 | 0.091 | 0.062 | 4 | 0.683 | 0.991 | 0.167 | 0.063 |
| 14 | 0.071 | 0.082 | 5 | 1.153 | 1.126 | 0.179 | 0.083 |
| 17 | 0.059 | 0.055 | 6 | 0.943 | 1.153 | 0.179 | 0.056 |

En este ejemplo: $\hat{s} = 3$ en ambos casos por lo que $\hat{a} = \dfrac{\hat{f}_k(\theta_4)}{f(\theta_4)} = 19.883$

y $\hat{b} = \dfrac{\hat{f}(\theta_4)}{f(\theta_4)} = 0.991$.

Parece que $\hat{b}$ pudiera estar estimando la constante 1 y en ese caso $\hat{f}(\theta)$ seria un estimador de $f(\theta) = \theta^{-1}$ y el calculo de $\hat{b}$ seria innecesario. Sin embargo dado que el error en la estimación de $\hat{a}$ afecta al estimador $\dfrac{\hat{f}_k(\theta)}{\hat{a}}$, se considera apropiado siempre comparar este estimador con $\dfrac{\hat{f}(\theta)}{\hat{b}}$ aunque se sospeche que la constante b sea igual a 1, de forma que el error proveniente del calculo de la constante afecte a ambos estimadores.

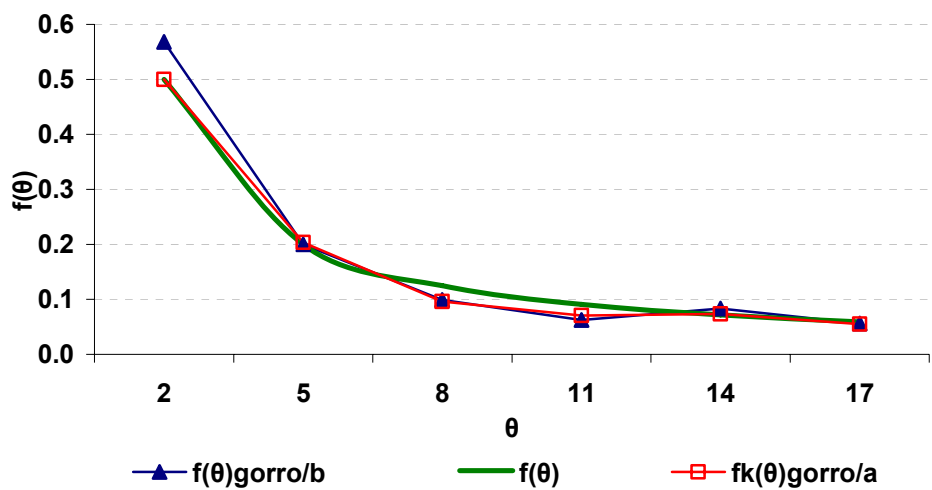

**Figura 2.2: Estimación de** $\dfrac{f_k(\theta)}{a}$ **y** $\dfrac{f(\theta)}{b}$ **para datos distribuidos** $Unif(0,\theta)$

Ya que tenemos 2 estimadores, $\dfrac{\hat{f}_k(\theta)}{\hat{a}}$ y $\dfrac{\hat{f}(\theta)}{\hat{b}}$ que estiman la misma función inicial de referencia (en este caso $f(\theta)=\theta^{-1}$) estamos en condiciones de comparar su desempeño.

Para comparar la precisión de los estimadores para varios valores de $\theta$ se calculara el Error Absoluto Relativo Promedio (EARP).

Retomando el ejemplo anterior se tienen los siguientes EARPs para ambos estimadores:

**Tabla 2.6: Cálculo del EARP para** $\dfrac{\hat{f}_k(\theta)}{\hat{a}}$ **y** $\dfrac{\hat{f}(\theta)}{\hat{b}}$ **para datos distribuidos** $Unif(0,\theta)$

| θ | f(θ) | f(θ)gorro | $f_k(\theta)$gorro | $f_k(\theta)$gorro/a | f(θ)gorro/b | EAR f(θ)gorro/b | EAR $f_k(\theta)$gorro/a |
|---|---|---|---|---|---|---|---|
| 2 | 0.500 | 0.563 | 9.941 | 0.500 | 0.568 | 0.137 | 0.000 |
| 5 | 0.200 | 0.198 | 4.047 | 0.204 | 0.200 | 0.000 | 0.018 |
| 8 | 0.125 | 0.098 | 1.907 | 0.096 | 0.099 | 0.209 | 0.233 |
| 11 | 0.091 | 0.062 | 1.397 | 0.070 | 0.063 | 0.311 | 0.227 |
| 14 | 0.071 | 0.082 | 1.455 | 0.073 | 0.083 | 0.164 | 0.024 |
| 17 | 0.059 | 0.055 | 1.095 | 0.055 | 0.056 | 0.048 | 0.063 |
| | | | | | EARP | 0.145 | 0.094 |

De donde se apoya la sospecha inicial que el estimador $\hat{f}(\theta)$ será menos preciso en general que el estimador $\hat{f}_k(\theta)$ debido a que tiene dos fuentes de variabilidad (el numerador y el denominador).

Se usarán tres tipos de distribución para generar las muestras aleatorias con el fin de evaluar empíricamente el error de las distintas variantes del método de estimación por simulación:

1. La Exponencial con esperanza $1/\theta$ ; denotada $p(y|\theta) \sim Exp(\theta)$

2. La Uniforme entre $\theta$ y $\theta^2$ ; denotada $p(y|\theta) \sim Unif(\theta, \theta^2)$

3. La Triangular con parámetros 0, $\theta$ y 1 ; denotada $p(y|\theta) \sim Trian(0, \theta, 1)$

Se escogió probar el desempeño de los estimadores para datos con distribución exponencial y triangular por su relevancia y aplicación dentro de la Ingeniería Industrial (la triangular en la estimación de duración de actividades en la Planeación de Proyectos y la exponencial en la estimación de tiempo entre llegadas en la Teoría de Colas). Además, para datos con distribución triangular el método por simulación es particularmente relevante dado que la derivación analítica de la inicial de referencia no parece ser factible [Ver Berger et al., 2009, p. 20]. La distribución uniforme entre $\theta$ y $\theta^2$ se escogió porque a diferencia de las otras dos, su soporte (los valores para los cuales la función de densidad es positiva) depende del parámetro.

## 2.2 Ejemplos y Resultados

### 2.2.1 Distribución Exponencial

El modelo está definido por:

$$p(y|\theta) = \theta e^{-\theta y}, 0 < y < \infty, 0 < \theta < \infty$$

$$p(x_j^{(k)}|\theta) = \prod_{i=1}^{k} p(y_{ij}|\theta) = \theta^k e^{-\theta \sum_{i=1}^{k} y_{ij}}$$

$$c_j = \int_{\Theta} p(x_j^{(k)}|\theta) d\theta = \int_0^\infty \theta^k e^{-\theta \sum_{i=1}^{k} y_{ij}} d\theta = \frac{k!}{\left(\sum_{i=1}^{k} y_{ij}\right)^{k+1}}$$

$$r_j(\theta) = \log(p(x_j^{(k)}|\theta)/c_j) = \log\left(\frac{\theta^k \left(\sum_{i=1}^{k} y_{ij}\right)^{k+1}}{e^{\theta \sum_{i=1}^{k} y_{ij}} k!}\right) = k\log(\theta) + (k+1)\log\left(\sum_{i=1}^{k} y_{ij}\right) - \theta \sum_{i=1}^{k} y_{ij} - (\log(k!))$$

Se sabe que la inicial de referencia para datos distribuidos exponencialmente, según el método de Bernardo es proporcional a $\theta^{-1}$ [Bernardo, 2005].

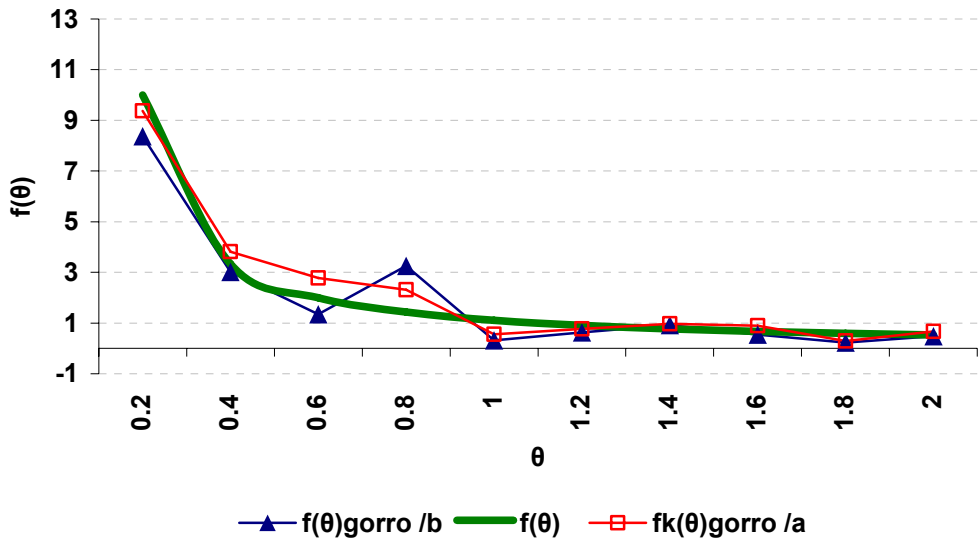

**Figura 2.3:** Estimación de $\dfrac{f_k(\theta)}{a}$ y $\dfrac{f(\theta)}{b}$ **para datos distribuidos** $Exp(\theta)$

En la figura 2.3 se muestran los resultados de la estimación de $f(\theta)$ para 10 diferentes valores de $\theta$ simulando 5 muestras aleatorias de tamaño 5 (k=5). Se muestran 2 diferentes estimadores de $f(\theta)$: $\dfrac{\hat{f}(\theta)}{\hat{b}}$ y $\dfrac{\hat{f}_k(\theta)}{\hat{a}}$

**Tabla 2.7: Comparación del EARP entre** $\dfrac{\hat{f}_k(\theta)}{\hat{a}}$ **y** $\dfrac{\hat{f}(\theta)}{\hat{b}}$ **para datos distribuidos** $Exp(\theta)$

|  | Repetición | | | | | | | | | |
| --- | --- | --- | --- | --- | --- | --- | --- | --- | --- | --- |
|  | 1 | 2 | 3 | 4 | 5 | 6 | 7 | 8 | 9 | 10 |
| $\dfrac{\hat{f}(\theta)}{\hat{b}}$ | 0.455 | 0.448 | 0.448 | 0.416 | 0.465 | 0.490 | 0.439 | 0.530 | 0.447 | 0.454 |
| $\dfrac{\hat{f}_k(\theta)}{\hat{a}}$ | 0.258 | 0.225 | 0.262 | 0.268 | 0.277 | 0.299 | 0.295 | 0.300 | 0.290 | 0.291 |

En la tabla 2.7 se compara el error absoluto relativo promedio entre los estimadores $\dfrac{\hat{f}_k(\theta)}{\hat{a}}$ y $\dfrac{\hat{f}(\theta)}{\hat{b}}$ para 10 diferentes simulaciones con k = 5. La tabla corrobora que, para este ejemplo, el estimador $\dfrac{\hat{f}_k(\theta)}{\hat{a}}$ es más preciso.

**2.2.2 Distribución Uniforme** $Unif(\theta, \theta^2)$

El modelo está definido por:

$$p(y|\theta) = \frac{1}{\theta(\theta-1)}, 1 < \theta < y < \theta^2$$

$$p(x_j^{(k)}|\theta) = \prod_{i=1}^{k} p(y_{ij}|\theta) = \frac{1}{\theta^k(\theta-1)^k} \prod_{i=1}^{k} I_{(\theta,\theta^2)}(y_{ij}) = \frac{1}{\theta^k(\theta-1)^k} I_{(\sqrt{t_{(k)j}}, t_{(1)j})}(\theta)$$

con $t_{(k)j} = \max_{i=1..k}\{y_{ij}\}, t_{(1)j} = \min_{i=1..k}\{y_{ij}\}$

$$c_j = \int_\Theta p(x_j^{(k)}|\theta)d\theta = \int_{\sqrt{t_{(k)j}}}^{t_{(1)j}} \frac{d\theta}{\theta^k(\theta-1)^k}$$

$$r_j(\theta) = \log(p(x_j^{(k)}|\theta)/c_j) = \log\left(\frac{1}{\left(\int_{\sqrt{t_{(k)j}}}^{t_{(1)j}} \frac{d\theta}{\theta^k(\theta-1)^k}\right)\theta^k(\theta-1)^k}\right)$$

$$= -k\log(\theta) - k\log(\theta-1) - \log\left(\int_{\sqrt{t_{(k)j}}}^{t_{(1)j}} \frac{d\theta}{\theta^k(\theta-1)^k}\right)$$

Se prueba que la inicial de referencia para datos con distribución $Unif(\theta, \theta^2)$ según el método de Bernardo es proporcional a $\dfrac{2\theta-1}{\theta(\theta-1)} e^{\psi\left(\frac{2\theta}{2\theta-1}\right)-1}, \psi = digamma$ [Berger et al., 2009].

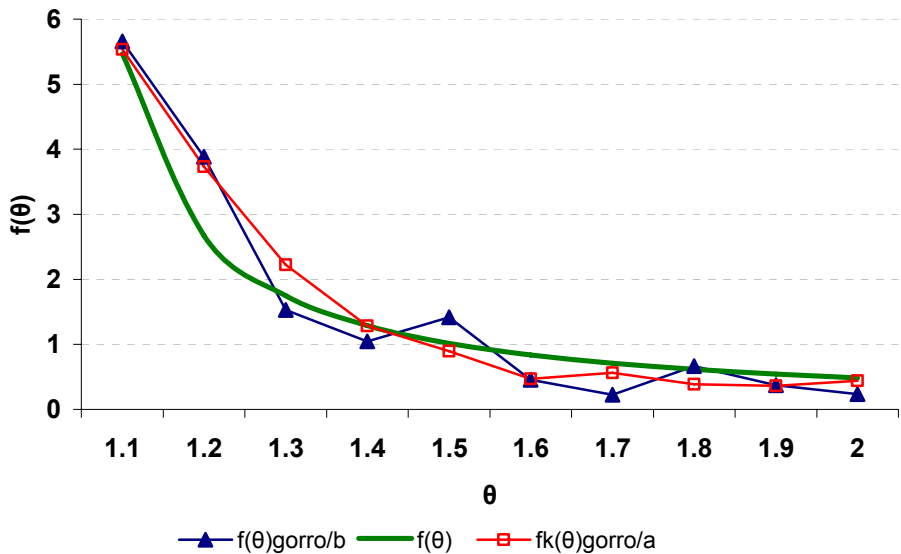

**Figura 2.4:** Estimación de $\dfrac{f_k(\theta)}{a}$ y $\dfrac{f(\theta)}{b}$ para datos distribuidos $Unif(\theta,\theta^2)$

En la figura 2.4 se muestran los resultados de la estimación de $f(\theta)$ para 10 diferentes valores de $\theta$ simulando 5 muestras aleatorias de tamaño 5 (k=5). Se muestran 2 diferentes estimadores de $f(\theta)$: $\dfrac{\hat{f}_k(\theta)}{\hat{a}}$ y $\dfrac{\hat{f}(\theta)}{\hat{b}}$.

**Tabla 2.8:** Comparación del EARP entre $\dfrac{\hat{f}_k(\theta)}{\hat{a}}$ y $\dfrac{\hat{f}(\theta)}{\hat{b}}$ para datos distribuidos $Unif(\theta,\theta^2)$

|  | Repetición | | | | | | | | | |
|---|---|---|---|---|---|---|---|---|---|---|
|  | 1 | 2 | 3 | 4 | 5 | 6 | 7 | 8 | 9 | 10 |
| $\dfrac{\hat{f}(\theta)}{\hat{b}}$ | 0.35 | 0.37 | 0.34 | 0.37 | 0.34 | 0.34 | 0.36 | 0.33 | 0.37 | 0.33 |
| $\dfrac{\hat{f}_k(\theta)}{\hat{a}}$ | 0.24 | 0.27 | 0.25 | 0.30 | 0.26 | 0.26 | 0.25 | 0.25 | 0.24 | 0.27 |

La tabla 2.8 muestra que, para este ejemplo, el estimador $\dfrac{\hat{f}_k(\theta)}{\hat{a}}$ es más preciso que el estimador $\dfrac{\hat{f}(\theta)}{\hat{b}}$. Se uso un valor de k = 5 para las 10 simulaciones.

### 2.2.3 Distribución Triangular

El modelo está definido por:

$$p(y|\theta) = \frac{2y}{\theta} I_{(0,\theta)}(y) + \frac{2(1-y)}{1-\theta} I_{(\theta,1)}(y)$$

$$p(x_j^{(k)}|\theta) = \prod_{i=1}^{k} p(y_{ij}|\theta) = 2^k \left(\prod_{i=1}^{r} t_{(i)j}\right)\left(\prod_{i=r+1}^{k} 1 - t_{(i)j}\right)\left(\frac{1}{\theta}\right)^r \left(\frac{1}{1-\theta}\right)^{k-r} \text{ con } t_{(r)j} < \theta < t_{(r+1)j}$$

$t_{(r)j}$ es la r-esima estadística de orden para $r \in \{1,...,k\}$ y $t_{(0)j} = 0, t_{(k+1)j} = 1$

$$c_j = 2^k \begin{bmatrix} \left(\prod_{i=1}^{k} 1 - t_{(i)j}\right)\left(\dfrac{(1-t_{(1)j})^{1-k}-1}{k-1}\right) + \left(\prod_{i=1}^{1} t_{(i)j}\right)\left(\prod_{i=2}^{k} 1 - t_{(i)j}\right)\int_{t_{(1)j}}^{t_{(2)j}} \left(\dfrac{1}{\theta}\right)^1 \left(\dfrac{1}{1-\theta}\right)^{k-1} d\theta \\ + ... + \left(\prod_{i=1}^{q} t_{(i)j}\right)\left(\prod_{i=q+1}^{k} 1 - t_{(i)j}\right)\int_{t_{(q)j}}^{t_{(q+1)j}} \left(\dfrac{1}{\theta}\right)^q \left(\dfrac{1}{1-\theta}\right)^{k-(q+1)} d\theta + ... + \\ \left(\prod_{i=1}^{k-1} t_{(i)j}\right)\left(\prod_{i=k}^{k} 1 - t_{(i)j}\right)\int_{t_{(k-1)j}}^{t_{(k)j}} \left(\dfrac{1}{\theta}\right)^{k-1} \left(\dfrac{1}{1-\theta}\right)^{k-(k-1)} d\theta + \left(\prod_{i=1}^{k} t_{(i)j}\right)\left(\dfrac{(t_{(k)j})^{1-k}-1}{k-1}\right) \end{bmatrix}$$

$$r_j(\theta) = \log\left(p(x_j^{(k)}|\theta)/c_j\right)$$

Aunque la derivación analítica de la inicial de referencia para datos distribuidos $Trian(0,\theta,1)$ según el método de Bernardo no es sencilla existen argumentos para pensar que es proporcional a $\dfrac{1}{\theta^{0.5}(1-\theta)^{0.5}}$ [Berger et al., 2009]. Es decir que no está probado que esta es en realidad la inicial de referencia. Es en estos casos en los cuales es útil el método de simulación y es parte de la razón por la cual se escogió este ejemplo, sin embargo las medidas de precisión de los estimadores se basan en el supuesto de que $\frac{1}{\theta^{0.5}(1-\theta)^{0.5}}$ es en realidad la inicial de referencia verdadera.

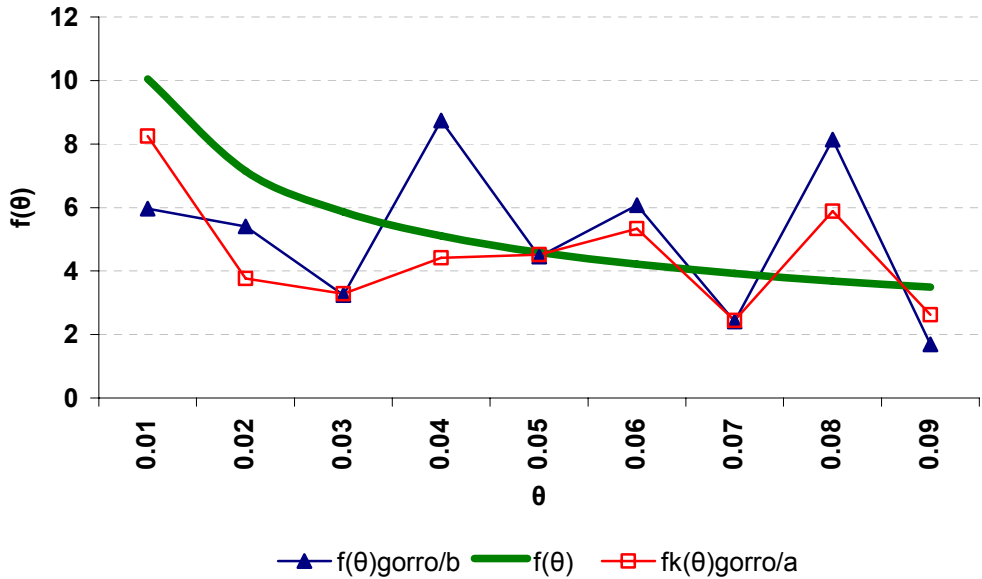

**Figura 2.5: Estimación de** $\frac{f_k(\theta)}{a}$ **y** $\frac{f(\theta)}{b}$ **para datos distribuidos** $Trian(0,\theta,1)$

En la Figura 2.5 se muestran los resultados de la estimación de $f(\theta)$ para 9 diferentes valores de $\theta$ simulando 5 muestras aleatorias de tamaño 5 (k=5). Se muestran 2 diferentes estimadores de $f(\theta)$: $\frac{\hat{f}_k(\theta)}{\hat{a}}$ y $\frac{\hat{f}(\theta)}{\hat{b}}$.

**Tabla 2.9: Comparación del EARP entre** $\frac{\hat{f}_k(\theta)}{\hat{a}}$ **y** $\frac{\hat{f}(\theta)}{\hat{b}}$ **para datos distribuidos** $Trian(0,\theta,1)$

|  | Repetición | | | | | | | | | |
|---|---|---|---|---|---|---|---|---|---|---|
|  | 1 | 2 | 3 | 4 | 5 | 6 | 7 | 8 | 9 | 10 |
| $\frac{\hat{f}(\theta)}{\hat{b}}$ | 0.329 | 0.296 | 0.286 | 0.300 | 0.303 | 0.266 | 0.281 | 0.287 | 0.342 | 0.303 |
| $\frac{\hat{f}_k(\theta)}{\hat{a}}$ | 0.251 | 0.217 | 0.206 | 0.234 | 0.264 | 0.219 | 0.231 | 0.238 | 0.265 | 0.212 |

La tabla 2.9 muestra que, para este ejemplo, el estimador $\dfrac{\hat{f}_k(\theta)}{\hat{a}}$ es más preciso que el estimador $\dfrac{\hat{f}(\theta)}{\hat{b}}$. Se uso un valor de k = 5 para las 10 simulaciones.

## 2.3 Conclusiones

En todos los ejemplos mostrados resultó más preciso el estimador $\dfrac{\hat{f}_k(\theta)}{\hat{a}}$ que $\dfrac{\hat{f}(\theta)}{\hat{b}}$. Esto era de esperarse pues en la estimación de $\hat{f}(\theta)$ se tienen que simular las k muestras dos veces: una vez para estimar $\hat{f}_k(\theta)$ y la otra para estimar $\hat{f}_k(\theta_0)$ y poder así calcular $\hat{f}(\theta) = \dfrac{\hat{f}_k(\theta)}{\hat{f}_k(\theta_0)}$.

Entonces, no solo es más preciso el estimador $\dfrac{\hat{f}_k(\theta)}{\hat{a}}$ sino que además requiere la mitad de simulaciones.

# 3. ESTIMACIÓN DE LA PRECISIÓN

## 3.1 Intervalos asintóticos de probabilidad para los Estimadores

Para determinar de qué tamaño debe ser la muestra obtenida por simulación (qué valor asignarle a *k*) para estimar $\hat{f}(\theta)$ y $\hat{f}_k(\theta)$ es necesario construir intervalos de probabilidad para $\hat{f}(\theta)$ y $\hat{f}_k(\theta)$ asintóticamente válidos (cuando *k* tiende a infinito) y analizar para qué tamaño de *k* se cumplen con la precisión deseada. Primero se determinará el intervalo asintótico de probabilidad de $\hat{f}(\theta)$ puesto que el intervalo asintótico de probabilidad de $\hat{f}_k(\theta)$ resultará ser un caso especial de éste.

Los resultados de esta sección se apoyan en dos teoremas conocidos en la literatura de procesos estocásticos, que enunciamos a continuación para facilitar la exposición. El símbolo "$\mapsto$" denota convergencia débil de variables aleatorias (cuando $k \to \infty$, a menos que se indique de otra forma)

**Teorema de Mapeo Continuo.** Para $k = 1,2,\ldots$, sean $X_k$ y $X$ variables aleatorias tomando valores en el espacio métrico $S$, y supóngase que $X_k \mapsto X$. Si $g : S \to \Re^p$ satisface que $P[X \in D(g)] = 0$ (donde $D(g)$ es el conjunto de puntos en los que $g$ es discontinua), entonces $g(X_k) \mapsto g(X)$.

**Teorema de Convergencia Conjunta.** Sean $\{X_k : k = 1,2,\ldots\}$ y $\{Y_k : k = 1,2,\ldots\}$ secuencias de variables aleatorias (con valores en $\Re^p$ y $\Re^q$, respectivamente). Supóngase que $X_k \mapsto X$, y $Y_k \mapsto a$ (donde $X$ es una variable aleatoria con valores en $\Re^p$, y $a \in \Re^q$ es una constante). Si $h : \Re^{p+q} \to \Re^r$ satisface $P[(X,a) \in D(h)] = 0$, entonces $h(X_k, Y_k) \mapsto h(X, a)$.

Una prueba del teorema de mapeo continuo puede encontrarse en el Corolario 2.1.9 de Ethier y Kurtz [1986] y una prueba del teorema de convergencia conjunta en Billingsley [1968].

### 3.1.1 Intervalo asintótico de probabilidad de $\hat{f}(\theta)$

Sean $\mu_1 = E_{p(x^{(k)}|\theta)}\left[\log\left[\dfrac{p(x^{(k)}|\theta)}{\int_\Theta p(x^{(k)}|\theta)d\theta}\right]\right]$, y $\mu_2 = E_{p(x^{(k)}|\theta_0)}\left[\log\left[\dfrac{p(x^{(k)}|\theta_0)}{\int_\Theta p(x^{(k)}|\theta)d\theta}\right]\right]$, entonces:

$$\hat{\mu}_1(k) = \frac{1}{k}\sum_{j=1}^{k}\log\left[\frac{p(x^{(k)}|\theta)}{\int_\Theta p(x^{(k)}|\theta)d\theta}\right] = \frac{1}{k}\sum_{j=1}^{k}r_j(\theta),$$

y

$$\hat{\mu}_2(k) = \frac{1}{k}\sum_{j=1}^{k}\log\left[\frac{p(x^{(k)}|\theta_0)}{\int_\Theta p(x^{(k)}|\theta)d\theta}\right] = \frac{1}{k}\sum_{j=1}^{k}r_j(\theta_0).$$

Si denotamos $\hat{\mu}(k) = [\hat{\mu}_1(k), \hat{\mu}_2(k)] = \dfrac{1}{k}\sum_{j=1}^{k}(r_j(\theta), r_j(\theta_0))$ y $\mu = (\mu_1, \mu_2)$, asumiendo momentos de segundo orden finitos para $r_j(\theta)$ y $r_j(\theta_0)$ se cumple un Teorema de Limite Central multivariado, ver e.g., Teorema B en la sección 1.91 de Serfling [1980]:

$$k^{1/2}[\hat{\mu}(k) - \mu] \mapsto GN_2(0, I),$$

donde $N_2(0, I)$ denota a una distribución normal (bivariada) con esperanza 0 y matriz varianza-covarianza $I$ (la identidad), $GG^T = \begin{bmatrix} \sigma_1^2 & \sigma_{12} \\ \sigma_{12} & \sigma_2^2 \end{bmatrix}$,

$\sigma_1^2 = V[r_j(\theta)] = E[r_j(\theta)^2] - (E[r_j(\theta)])^2$, $\sigma_2^2 = V[r_j(\theta_0)] = E[r_j(\theta_0)^2] - (E[r_j(\theta_0)])^2$ y

$\sigma_{12} = Cov[r_j(\theta)r_j(\theta_0)] = E[r_j(\theta)r_j(\theta_0)] - E[r_j(\theta)]E[r_j(\theta_0)]$. Nótese que este Teorema del Límite Central implica que:

$$\hat{\mu}(k) \mapsto [\mu_1(k), \mu_2(k)] = \mu(k).$$

Por otro lado, aplicando el método Delta, ver e.g., Proposición 2 de Muñoz y Glynn [1997], se tiene que $k^{1/2}[g(\hat{\mu}(k)) - g(\mu)] \mapsto \sigma N(0,1)$ donde $\sigma^2 = [\nabla g(\mu)]^T GG^T \nabla g(\mu)$. En este caso tomamos $g(\mu) = \dfrac{e^{\mu_1}}{e^{\mu_2}} = e^{\mu_1 - \mu_2} = f(\theta)$ y $g(\hat{\mu}) = \dfrac{e^{\hat{\mu}_1}}{e^{\hat{\mu}_2}} = e^{\hat{\mu}_1 - \hat{\mu}_2} = \hat{f}(\theta)$. De esta manera tenemos que $k^{1/2}[e^{\hat{\mu}_1 - \hat{\mu}_2} - e^{\mu_1 - \mu_2}] = k^{1/2}[\hat{f}(\theta) - f(\theta)] \mapsto \sigma N(0,1)$ con

$$\sigma^2 = [\nabla g(\mu)]^T GG^T \nabla g(\mu) = \begin{bmatrix} e^{\mu_1 - \mu_2} & -e^{\mu_1 - \mu_2} \end{bmatrix} \begin{bmatrix} \sigma_1^2 & \sigma_{12} \\ \sigma_{12} & \sigma_2^2 \end{bmatrix} \begin{bmatrix} e^{\mu_1 - \mu_2} \\ -e^{\mu_1 - \mu_2} \end{bmatrix}$$

$$= e^{2\mu_1 - 2\mu_2} (\sigma_1^2 + \sigma_2^2 - 2\sigma_{12})$$

En resumen tenemos que $\dfrac{k^{1/2}[e^{\hat{\mu}_1 - \hat{\mu}_2} - e^{\mu_1 - \mu_2}]}{\sigma} = \dfrac{k^{1/2}[\hat{f}(\theta) - f(\theta)]}{\sigma} \mapsto N(0,1)$.

Si escogemos $\hat{\sigma}_1^2$, $\hat{\sigma}_2^2$ y $\hat{\sigma}_{12}$, estimadores consistentes de $\sigma_1^2$, $\sigma_2^2$ y $\sigma_{12}$ respectivamente entonces tenemos que $\hat{\sigma} = e^{\hat{\mu}_1(k) - \hat{\mu}_2(k)} \sqrt{\hat{\sigma}_1^2 + \hat{\sigma}_2^2 - 2\hat{\sigma}_{12}}$ es un estimador consistente de $\sigma$ (teorema de mapeo continuo), se concluye del teorema de convergencia conjunta que:

$$\dfrac{k^{1/2}[e^{\hat{\mu}_1 - \hat{\mu}_2} - e^{\mu_1 - \mu_2}]}{\hat{\sigma}} = \dfrac{k^{1/2}[\hat{f}(\theta) - f(\theta)]}{\hat{\sigma}} \mapsto N(0,1)$$

Como los estimadores $\hat{\sigma}_1^2 = \dfrac{\sum_{j=1}^{k}(r_j(\theta) - \hat{\mu}_1)^2}{k-1} \to V[r_j(\theta)] = \sigma_1^2$ y

$\hat{\sigma}_2^2 = \dfrac{\sum_{j=1}^{k}(r_j(\theta_0) - \hat{\mu}_2)^2}{k-1} \to V[r_j(\theta_0)] = \sigma_2^2$ con probabilidad 1 si $k \to \infty$, estos estimadores son consistentes y se usaran en el cálculo de $\hat{\sigma} = e^{\hat{\mu}_1(k) - \hat{\mu}_2(k)} \sqrt{\hat{\sigma}_1^2 + \hat{\sigma}_2^2 - 2\hat{\sigma}_{12}}$.

[Ver Teorema A, Serfling, 1980, p.69]

Si definimos $Z_j = r_j(\theta) r_j(\theta_0)$ entonces $\dfrac{\sum_{j=1}^{k} Z_j}{k} \to E[Z_j]$ con probabilidad 1 si $k \to \infty$. En consecuencia

$$\hat{\sigma}_{12} = \dfrac{\left\{\left[\dfrac{\sum_{j=1}^{k} r_j(\theta) r_j(\theta_0)}{k}\right] - \left[\left(\dfrac{\sum_{j=1}^{k} r_j(\theta)}{k}\right)\left(\dfrac{\sum_{j=1}^{k} r_j(\theta_0)}{k}\right)\right]\right\}}{\left(\dfrac{k-1}{k}\right)} \to Cov[r_j(\theta) r_j(\theta_0)] = \sigma_{12}$$

con probabilidad 1 si $k \to \infty$, por lo que este estimador es consistente y se usara en el cálculo de

$$\hat{\sigma} = e^{\hat{\mu}_1(k) - \hat{\mu}_2(k)} \sqrt{\hat{\sigma}_1^{\,2} + \hat{\sigma}_2^{\,2} - 2\hat{\sigma}_{12}}\ .$$

[Ver Teorema A, Serfling, 1980, p.67]

Ahora suponer que se tiene un modelo $M = \{p(x \mid \lambda): x \in \aleph^k, \lambda \in \Lambda \subset \Re\}$ y sea $\hat{\lambda}$ un estimador de $\lambda$ con función de densidad simétrica, es decir que para toda $0 < \beta < 1$ exista $z$ tal que $P[\hat{\lambda} - z < \lambda < \hat{\lambda} + z] = \beta$. Se define el ancho medio de un estimador $\hat{\lambda}$ del parámetro $\lambda$ como $AM_\alpha(\hat{\lambda})$ tal que

$$P[\hat{\lambda} - AM_\alpha < \lambda < \hat{\lambda} + AM_\alpha] = 1 - \alpha$$

Igualmente, se define el límite superior $L_{\alpha, SUP}(\hat{\lambda})$ de un intervalo de probabilidad para $\lambda$ como $L_{\alpha, SUP} = \hat{\lambda} + AM_\alpha$ y el límite inferior $L_{\alpha, INF}(\hat{\lambda})$ de un intervalo de probabilidad para $\lambda$ como $L_{\alpha, INF} = \hat{\lambda} - AM_\alpha$.

Para obtener el ancho medio de $\hat{f}(\theta)$, $AM_\alpha\left(\hat{f}(\theta)\right)$ sólo queda por desarrollar el siguiente intervalo asintótico de probabilidad:

$$P\left[-z_{1-\alpha/2} < \frac{k^{1/2}\left[\hat{f}(\theta) - f(\theta)\right]}{\hat{\sigma}} < z_{1-\alpha/2}\right] = 1 - \alpha \text{ entonces,}$$

$$P\left[\hat{f}(\theta) - \frac{z_{1-\alpha/2}\hat{\sigma}}{k^{1/2}} < f(\theta) < \hat{f}(\theta) + \frac{z_{1-\alpha/2}\hat{\sigma}}{k^{1/2}}\right] = 1 - \alpha$$

Por lo tanto el ancho medio de $\hat{f}(\theta)$ es:

$$AM_\alpha\left(\hat{f}(\theta)\right) = \frac{z_{1-\alpha/2}\hat{\sigma}}{k^{1/2}} = \frac{z_{1-\alpha/2} e^{\hat{\mu}_1(k) - \hat{\mu}_2(k)}\sqrt{\hat{\sigma}_1^2 + \hat{\sigma}_2^2 - 2\hat{\sigma}_{12}}}{k^{1/2}}.$$

### 3.1.2 Intervalo asintótico de probabilidad de $\hat{f}_k(\theta)$

En este caso tomamos $g(\mu) = e^{\mu_1} = f_k(\theta)$ y $g(\hat{\mu}) = e^{\hat{\mu}_1} = \hat{f}_k(\theta)$. De esta manera tenemos que

$$\left[k^{1/2}\left[e^{\hat{\mu}_1} - e^{\mu_1}\right]\right] = k^{1/2}\left[\hat{f}_k(\theta) - f_k(\theta)\right] \mapsto \sigma N(0,1) \text{ con}$$

$$\sigma^2 = \left[\nabla g(\mu)\right]^T GG^T \nabla g(\mu) = \left[\frac{dg(\mu)}{d\mu}\right]^2 \sigma_1^2 = e^{2\mu_1}\sigma_1^2$$

En resumen tenemos que $\dfrac{k^{1/2}\left[e^{\hat{\mu}_1} - e^{\mu_1}\right]}{\sigma} = \dfrac{k^{1/2}\left[\hat{f}_k(\theta) - f(\theta)\right]}{\sigma} \mapsto N(0,1)$. Además como

$\hat{\sigma} = e^{\hat{\mu}_1(k)}\hat{\sigma}_1$ es un estimador consistente de $\sigma$ por el teorema de convergencia conjunta

tenemos que: $\dfrac{k^{1/2}\left[e^{\hat{\mu}_1} - e^{\mu_1}\right]}{\hat{\sigma}} = \dfrac{k^{1/2}\left[\hat{f}_k(\theta) - f(\theta)\right]}{\hat{\sigma}} \mapsto N(0,1)$.

Para obtener el ancho medio de $\hat{f}_k(\theta)$, $AM_\alpha\left(\hat{f}_k(\theta)\right)$ solo queda desarrollar el siguiente intervalo asintótico de probabilidad:

$$P\left[z_{1-\alpha/2} < \frac{k^{1/2}\left[\hat{f}_k(\theta) - f(\theta)\right]}{\hat{\sigma}} < -z_{1-\alpha/2}\right] = 1 - \alpha \Rightarrow P\left[\hat{f}_k(\theta) + \frac{z_{1-\alpha/2}\hat{\sigma}}{k^{1/2}} < f(\theta) < \hat{f}_k(\theta) - \frac{z_{1-\alpha/2}\hat{\sigma}}{k^{1/2}}\right] = 1 - \alpha$$

Por lo tanto el ancho medio de $\hat{f}_k(\theta)$ es: $AM_\alpha\left(\hat{f}_k(\theta)\right) = \dfrac{z_{1-\alpha/2}\hat{\sigma}}{k^{1/2}} = z_{1-\alpha/2} k^{-1/2} e^{\hat{\mu}_1(k)} \hat{\sigma}_1$

Siguiendo con el ejemplo a partir de datos con distribución uniforme entre 0 y $\theta$, se calculan los anchos medios y límites inferior y superior de ambos estimadores para $\theta$ = 2, 5, 8, 11, 14 y 17 con k=5, $\theta_0 = 1$ y $\alpha = 0.1$.

**Tabla 3.1: Cálculo de anchos medios y límites inferior y superior de $\dfrac{\hat{f}_k(\theta)}{\hat{a}}$ y $\dfrac{\hat{f}(\theta)}{\hat{b}}$ para datos distribuidos $Unif(0,\theta)$**

| $\theta$ | $f(\theta)$ | $f_k(\theta)$gorro/a | $f(\theta)$gorro/b | AM $[f_k(\theta)$gorro/a] | AM $[f(\theta)$gorro/b] | $L_{SUP}$ $f(\theta)$gorro/b | $L_{INF}$ $f(\theta)$gorro/b | $L_{SUP}$ $f_k(\theta)$gorro/a | $L_{INF}$ $f_k(\theta)$gorro/a |
|---|---|---|---|---|---|---|---|---|---|
| 2 | 0.500 | 0.103 | 0.184 | 0.133 | 0.394 | 0.577 | -0.210 | 0.235 | -0.030 |
| 5 | 0.200 | 0.229 | 0.165 | 0.049 | 0.027 | 0.193 | 0.138 | 0.277 | 0.180 |
| 8 | 0.125 | 0.115 | 0.176 | 0.042 | 0.139 | 0.315 | 0.038 | 0.157 | 0.073 |
| 11 | 0.091 | 0.036 | 0.075 | 0.047 | 0.082 | 0.158 | -0.007 | 0.083 | -0.011 |
| 14 | 0.071 | 0.043 | 0.055 | 0.030 | 0.056 | 0.111 | -0.001 | 0.073 | 0.012 |
| 17 | 0.059 | 0.040 | 0.064 | 0.032 | 0.110 | 0.174 | -0.046 | 0.072 | 0.008 |

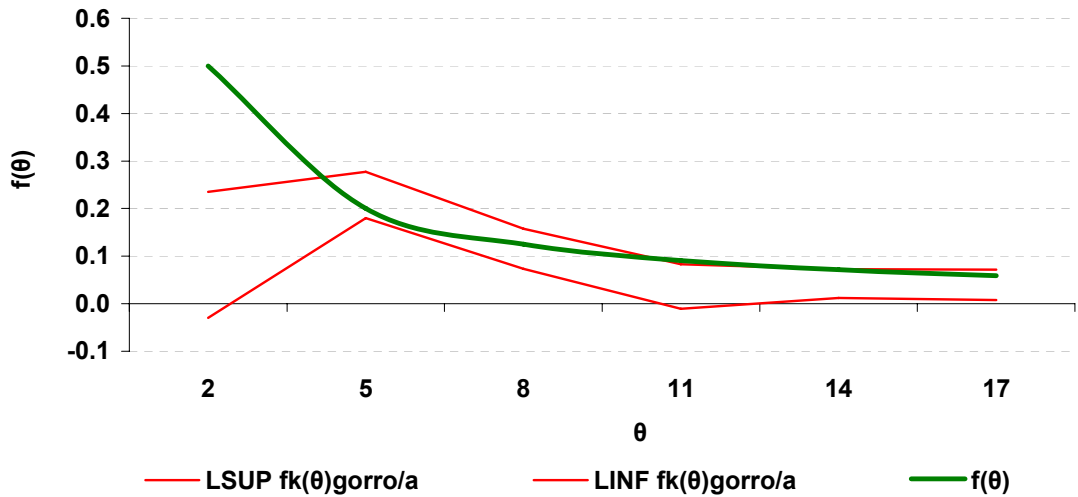

**Figura 3.1: Límites inferior y superior de de $\dfrac{\hat{f}_k(\theta)}{\hat{a}}$ para datos distribuidos $Unif(0,\theta)$**

Los intervalos de probabilidad y anchos medios construidos en esta sección se cumplen asintóticamente. En este ejemplo observar $f(\theta)$ no está incluido dentro del límite inferior y superior del estimador $\dfrac{\hat{f}_k(\theta)}{\hat{a}}$ en 2 ocasiones (para $\theta$ = 2 y 11) o en 2/6= 33.3% de los valores de $\theta$. Para k grande según el intervalo de probabilidad que construimos, podemos esperar que en únicamente 10% de los casos $f(\theta)$ caiga afuera de su límite inferior y superior, por lo que parece que k=5 no es suficientemente grande. Definimos la cobertura empírica de un procedimiento de estimación por intervalos de un parámetro $\lambda$, como el porcentaje de veces en las que $\lambda$ cae dentro de una colección de intervalos obtenidos con ese procedimiento a partir de datos simulados, es decir si se calculan los estimadores y límites inferior y superior para R valores de $\lambda$ la cobertura empírica $CE_\alpha$ se calcula como:

$$CE(\hat{\lambda}) = \sum_{l=1}^{R} \dfrac{I_{(L_{\alpha,INF,}(\hat{\lambda}_l) L_{\alpha,SUP}(\hat{\lambda}_l))}(\lambda_l)}{R},$$

Quisiéramos determinar para que valores de k podemos esperar que la cobertura empírica se aproxime a la probabilidad $1-\alpha$ con la que se construyó el intervalo. Regresando al ejemplo a partir de datos con distribución uniforme entre 0 y $\theta$, se calcula la cobertura empírica de ambos estimadores para 300 distintos valores de $\theta$ entre 2 y 899 y para k = 2,3,…,100 con los siguientes resultados mostrados gráficamente.

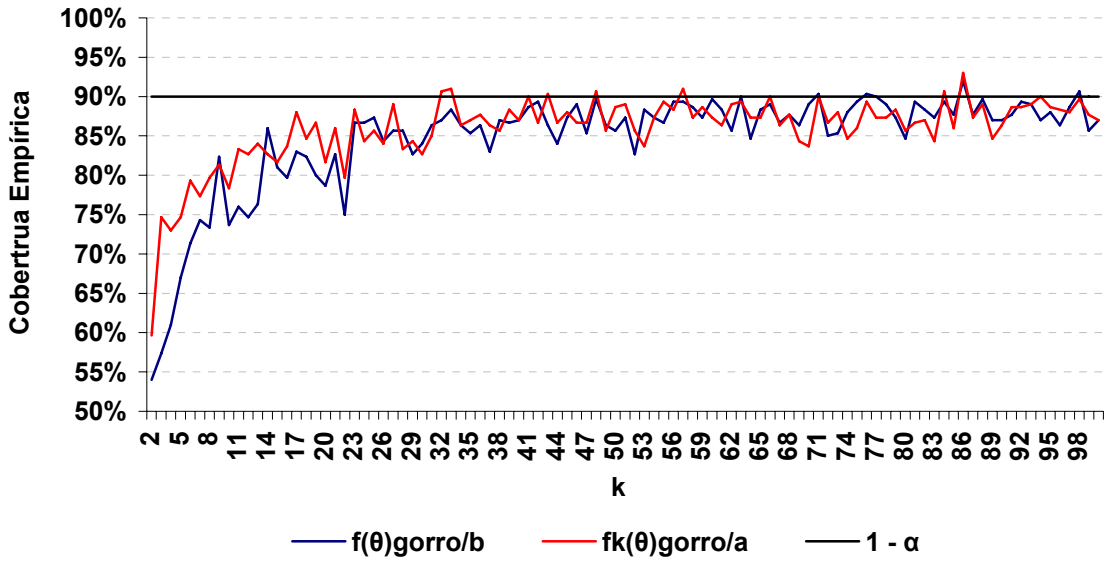

**Figura 3.2: Cobertura empírica de** $\dfrac{\hat{f}_k(\theta)}{\hat{a}}$ **y** $\dfrac{\hat{f}(\theta)}{\hat{b}}$ **para datos distribuidos** $Unif(0,\theta)$

Parece, que para este ejemplo ambos estimadores logran coberturas empíricas muy cercanas a las esperadas, según el intervalo asintótico de probabilidad, para valores de k = 50 o mayores, aunque parece que para $k < 20$  $\dfrac{\hat{f}_k(\theta)}{\hat{a}}$ obtiene coberturas empíricas más altas.

Para comparar la precisión de los estimadores para varios valores de $\theta$ se calculara el Ancho Medio Relativo Promedio (AMRP). Si se calculan los estimadores y sus anchos medios para R distintos valores de $\theta$ este se calcula como:

$$AMRP\left(\dfrac{\hat{f}(\theta)}{\hat{b}}\right) = \sum_{l=1}^{R} \dfrac{AM_\alpha\left(\hat{f}(\theta_l)\right)}{bf(\theta_l)R} \text{ y } AMRP\left(\dfrac{\hat{f}_k(\theta)}{\hat{a}}\right) = \sum_{l=1}^{R} \dfrac{AM_\alpha\left(\hat{f}_k(\theta_l)\right)}{af(\theta_l)R}$$

Notar que como en el caso del EARP la cobertura empírica y el AMRP evalúan el desempeño de los estimadores en varios valores de $\theta$ símultáneamente.

Siguiendo con el ejemplo a partir de datos con distribución uniforme entre $0$ y $\theta$, se calcula el AMRP de ambos estimadores para $\theta = 2, 5, 8, 11, 14$ y $17$ con k=5, $\theta_0 = 1$ y $\alpha = 0.1)$.

**Tabla 3.2: Cálculo de AMRP de** $\dfrac{\hat{f}_k(\theta)}{\hat{a}}$ **y** $\dfrac{\hat{f}(\theta)}{\hat{b}}$ **para datos distribuidos** $Unif(0,\theta)$

| θ | f(θ) | f(θ)gorro | $f_k(\theta)$gorro | AM [f(θ)gorro] | AM [$f_k(\theta)$gorro] | Ancho Medio Relativo f(θ)gorro/b | Ancho Medio Relativo $f_k(\theta)$gorro/a |
|---|------|-----------|---------------------|----------------|--------------------------|-----------------------------------|---------------------------------------------|
| 2  | 0.500 | 0.563 | 9.941 | 0.129 | 1.755 | 0.249 | 0.187 |
| 5  | 0.200 | 0.198 | 4.047 | 0.052 | 0.742 | 0.252 | 0.197 |
| 8  | 0.125 | 0.098 | 1.907 | 0.035 | 0.434 | 0.272 | 0.184 |
| 11 | 0.091 | 0.062 | 1.397 | 0.023 | 0.418 | 0.248 | 0.244 |
| 14 | 0.071 | 0.082 | 1.455 | 0.023 | 0.290 | 0.305 | 0.216 |
| 17 | 0.059 | 0.055 | 1.095 | 0.022 | 0.312 | 0.359 | 0.282 |
|    |       |       |       |       | **Promedio** | **0.281** | **0.218** |

Para este ejemplo el estimador $\dfrac{\hat{f}_k(\theta)}{\hat{a}}$ logra, en promedio, anchos medios relativos menores a los del estimador $\dfrac{\hat{f}(\theta)}{\hat{b}}$ sugiriendo una mayor precisión.

En el resto del capítulo 3 y en el 4 se usaran 3 criterios conjuntamente para evaluar el desempeño de los estimadores:

1. Ancho Medio Relativo Promedio (AMRP): Con esta medida se evaluará la precisión asintótica teórica de los estimadores de acuerdo a los intervalos de probabilidad construidos para los estimadores.

2. Cobertura empírica (CE): Con esta medida se evaluará en que grado y para que tamaños de muestra se cumple la precisión teórica medida con el AMRP.

3. Error Absoluto Relativo Promedio (EARP): Esta medida evalúa la precisión observada de los estimadotes.

## 3.2 Ejemplos y Resultados

### 3.2.1 Distribución Exponencial

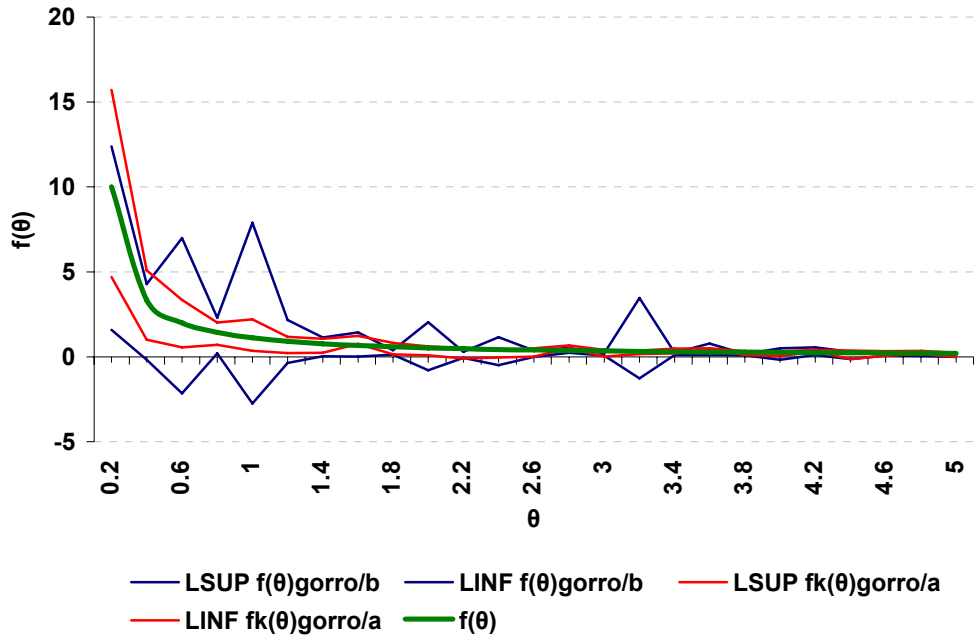

**Figura 3.3: Límites inferior y superior de** $\dfrac{\hat{f}_k(\theta)}{\hat{a}}$ **y** $\dfrac{\hat{f}(\theta)}{\hat{b}}$ **para datos distribuidos** $Exp(\theta)$

En la figura 3.3 se muestran los límites inferior y superior para una simulación con k = 5 y valor de $\alpha$ de 0.05. Parece que en general los límites del estimador $\dfrac{\hat{f}_k(\theta)}{\hat{a}}$ conforman intervalos más angostos por lo que se puede esperar mayor precisión de este estimador.

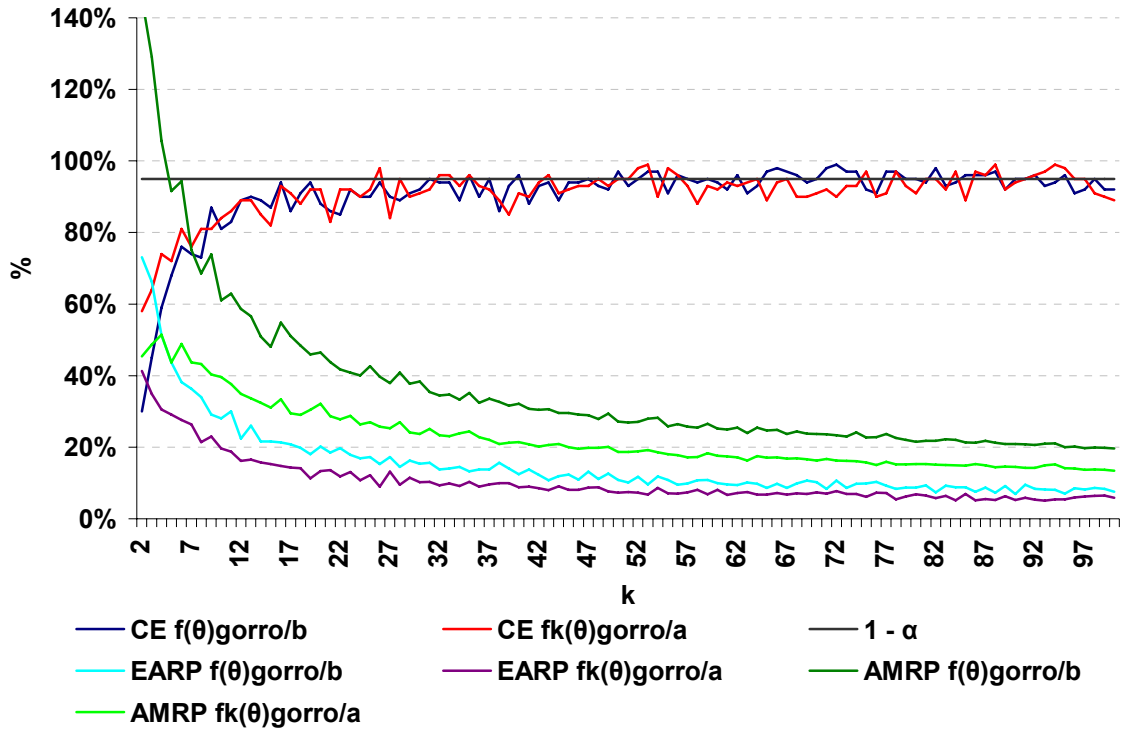

**Figura 3.4: CE, AMRP y EARP de $\dfrac{\hat{f}_k(\theta)}{\hat{a}}$ y $\dfrac{\hat{f}(\theta)}{\hat{b}}$ para datos distribuidos $Exp(\theta)$**

La figura 3.4 muestra la cobertura empírica, ancho medio relativo promedio y error absoluto relativo promedio de los estimadores $\dfrac{\hat{f}_k(\theta)}{\hat{a}}$ y $\dfrac{\hat{f}(\theta)}{\hat{b}}$ conforme k crece. Para cada valor de k se simularon 100 valores de $\theta$. Los límites inferior y superior se construyeron con un valor de alfa igual a 0.05. Se muestra que para tamaños de muestra mayores a 30 ambos estimadores logran coberturas empíricas similares $1-\alpha$. El AMRP del estimador $\dfrac{\hat{f}_k(\theta)}{\hat{a}}$ es siempre menor al de $\dfrac{\hat{f}(\theta)}{\hat{b}}$ y como las cobertura empírica de $\dfrac{\hat{f}_k(\theta)}{\hat{a}}$ es mayor a la de $\dfrac{\hat{f}(\theta)}{\hat{b}}$ para tamaños de muestra menores a 10 y muy similar para tamaños de muestra mayores el EARP de $\dfrac{\hat{f}_k(\theta)}{\hat{a}}$ es menor al de $\dfrac{\hat{f}(\theta)}{\hat{b}}$. Visto de otra forma se necesita un tamaño de muestra de alrededor de 35 para obtener un EARP de 10% con $\dfrac{\hat{f}_k(\theta)}{\hat{a}}$ y de alrededor de 60 para obtener un EARP similar con $\dfrac{\hat{f}(\theta)}{\hat{b}}$.

### 3.2.2 Distribución Uniforme ($\theta$, $\theta^2$)

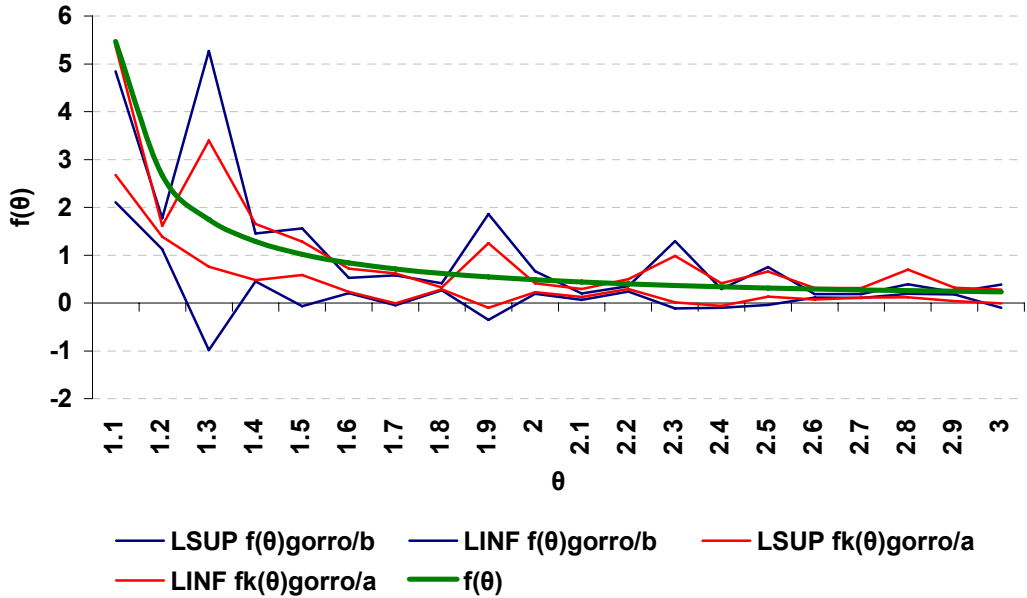

**Figura 3.5:** Límites inferior y superior de $\dfrac{\hat{f}_k(\theta)}{\hat{a}}$ y $\dfrac{\hat{f}(\theta)}{\hat{b}}$ **para datos distribuidos** $Unif(\theta, \theta^2)$

En la figura 3.5 se muestran los límites inferior y superior para una simulación con k = 3 y valor de $\alpha$ de 0.1. Parece que en general los límites del estimador $\dfrac{\hat{f}_k(\theta)}{\hat{a}}$ conforman intervalos más angostos por lo que se puede esperar mayor precisión de este estimador.

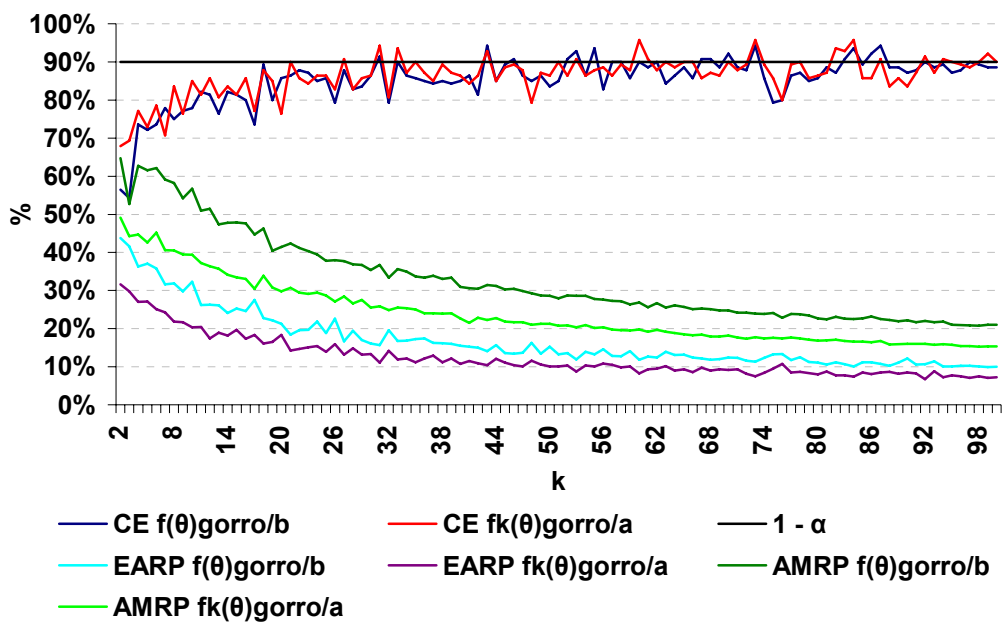

**Figura 3.6: CE, AMRP y EARP de** $\dfrac{\hat{f}_k(\theta)}{\hat{a}}$ **y** $\dfrac{\hat{f}(\theta)}{\hat{b}}$ **para datos distribuidos** $Unif(\theta,\theta^2)$

La figura 3.6 muestra la cobertura empírica, ancho medio relativo promedio y error absoluto relativo promedio de los estimadores $\dfrac{\hat{f}_k(\theta)}{\hat{a}}$ y $\dfrac{\hat{f}(\theta)}{\hat{b}}$ conforme k crece. Para cada valor de k se simularon 140 valores de $\theta$. Los límites inferior y superior se construyeron con un valor de alfa igual a 0.1. Se muestra que para tamaños de muestra mayores a 35 ambos estimadores logran coberturas empíricas similares $1-\alpha$. El AMRP del estimador $\dfrac{\hat{f}_k(\theta)}{\hat{a}}$ es siempre menor al de $\dfrac{\hat{f}(\theta)}{\hat{b}}$ y como las cobertura empírica de $\dfrac{\hat{f}_k(\theta)}{\hat{a}}$ es mayor a la de $\dfrac{\hat{f}(\theta)}{\hat{b}}$ para tamaños de muestra menores a 10 y muy similar para tamaños de muestra mayores el EARP de $\dfrac{\hat{f}_k(\theta)}{\hat{a}}$ es menor al de $\dfrac{\hat{f}(\theta)}{\hat{b}}$. Visto de otra forma se necesita un tamaño de muestra de alrededor de 60 para obtener un EARP de 10% con $\dfrac{\hat{f}_k(\theta)}{\hat{a}}$ y de alrededor de 100 para obtener un EARP similar con $\dfrac{\hat{f}(\theta)}{\hat{b}}$.

### 3.2.3 Distribución Triangular

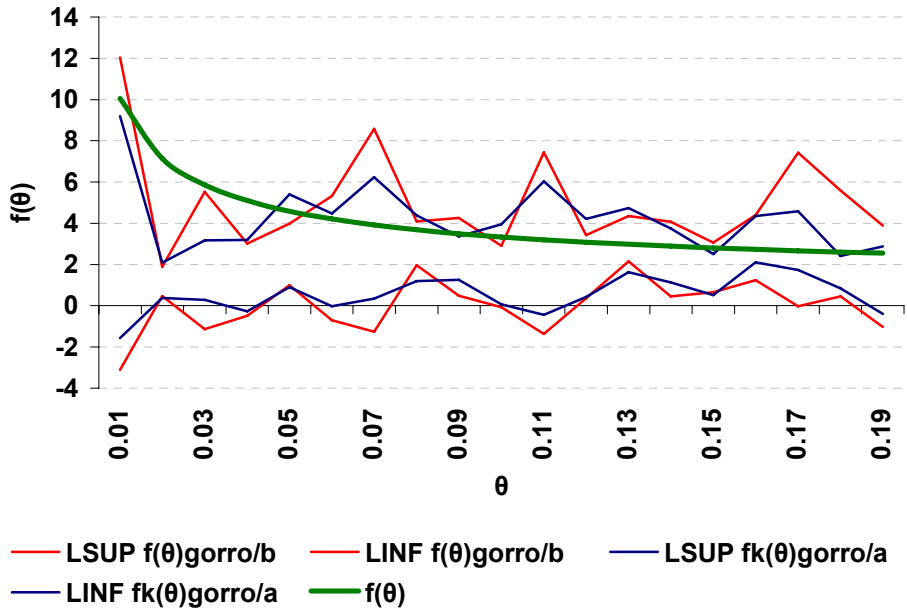

**Figura 3.7: Límites inferior y superior de $\dfrac{\hat{f}_k(\theta)}{\hat{a}}$ y $\dfrac{\hat{f}(\theta)}{\hat{b}}$ para datos distribuidos** $Trian(0,\theta,1)$

En la figura 3.7 se muestran las bandas de ancho para una simulación con k = 5 y valor de $\alpha$ de 0.08. En este caso no se puede apreciar una diferencia evidente entre el grosor de los intervalos conformados por los límites de $\dfrac{\hat{f}_k(\theta)}{\hat{a}}$ y los de $\dfrac{\hat{f}(\theta)}{\hat{b}}$ por lo que es necesario observar los errores absolutos relativos y anchos medios relativos para discernir si existe alguna diferencia con respecto a la precisión de ambos estimadores.

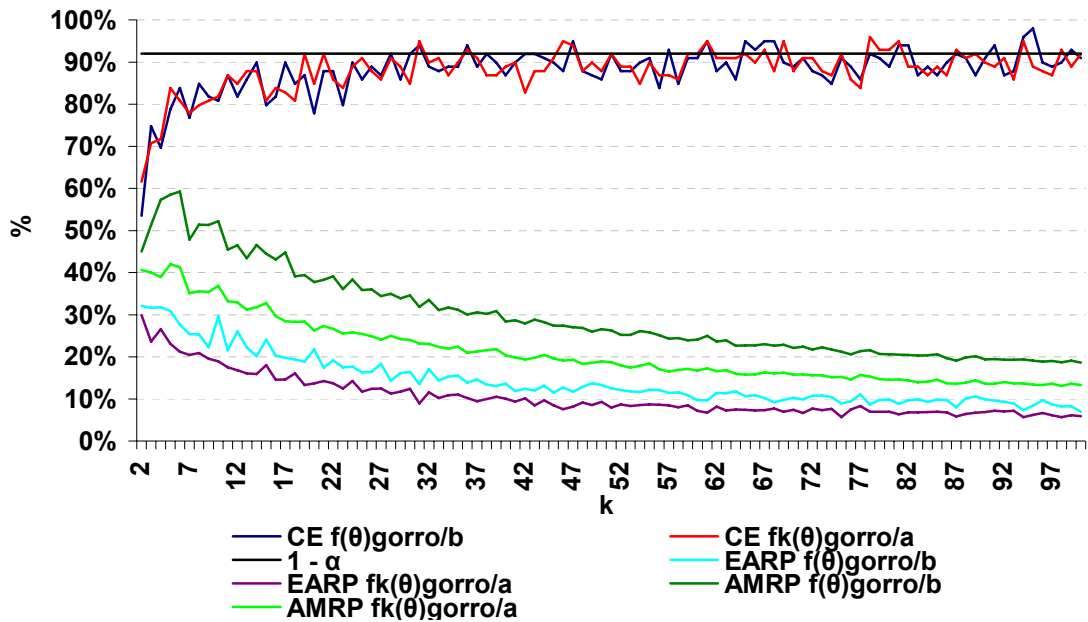

**Figura 3.8:** CE, AMRP y EARP de $\dfrac{\hat{f}_k(\theta)}{\hat{a}}$ y $\dfrac{\hat{f}(\theta)}{\hat{b}}$ **para datos distribuidos** $Trian(0,\theta,1)$

La figura 3.8 muestra la cobertura empírica, ancho medio relativo promedio y error absoluto relativo promedio de los estimadores $\dfrac{\hat{f}_k(\theta)}{\hat{a}}$ y $\dfrac{\hat{f}(\theta)}{\hat{b}}$ conforme k crece. Para cada valor de k se simularon 99 valores de $\theta$. Los límites inferior y superior se construyeron con un valor de alfa igual a 0.08. Se muestra que para tamaños de muestra mayores a 35 ambos estimadores logran coberturas empíricas similares $1-\alpha$. El AMRP del estimador $\dfrac{\hat{f}_k(\theta)}{\hat{a}}$ es siempre menor al de $\dfrac{\hat{f}(\theta)}{\hat{b}}$ y como las cobertura empírica de $\dfrac{\hat{f}_k(\theta)}{\hat{a}}$ es siempre muy similar a la de $\dfrac{\hat{f}(\theta)}{\hat{b}}$ el EARP de $\dfrac{\hat{f}_k(\theta)}{\hat{a}}$ es menor al de $\dfrac{\hat{f}(\theta)}{\hat{b}}$. Visto de otra forma se necesita un tamaño de muestra de alrededor de 40 para obtener un EARP de 10% con $\dfrac{\hat{f}_k(\theta)}{\hat{a}}$ y de alrededor de 70 para obtener un EARP similar con $\dfrac{\hat{f}(\theta)}{\hat{b}}$.

**3.3 Conclusiones**

Para un mismo valor de $\alpha$ y de $k$ tenemos que, para los tres ejemplos estudiados, el ancho medio de $\dfrac{\hat{f}(\theta)}{\hat{b}}$ es mayor al de $\dfrac{\hat{f}_k(\theta)}{\hat{a}}$ lo que, en conjunto con coberturas empíricas similares para los dos estimadores, explica porque el estimador $\dfrac{\hat{f}_k}{\hat{a}}(\theta)$ es más preciso que $\dfrac{\hat{f}(\theta)}{\hat{b}}$.

En los tres ejemplos estudiados se necesitó una valor de $k$ de alrededor de 30 para obtener coberturas empíricas muy similares a $1-\alpha$. Además se observó que en dos de los tres ejemplos se obtienen coberturas empíricas mayores con $\dfrac{\hat{f}_k(\theta)}{\hat{a}}$ para valores pequeños de $k$ (menores a 10). Si además se considera que el estimador $\dfrac{\hat{f}(\theta)}{\hat{b}}$ requiere la simulación del doble de números aleatorios que $\dfrac{\hat{f}_k(\theta)}{\hat{a}}$ resulta que este último tiene mejor desempeño que $\dfrac{\hat{f}(\theta)}{\hat{a}}$ en todos los sentidos considerados.

# 4. NÚMEROS ALEATORIOS COMUNES

## 4.1 El estimador de Números Aleatorios Comunes (NAC)

Para calcular el estimador $\hat{f}(\theta) = \dfrac{\hat{f}_k(\theta)}{\hat{f}_k(\theta_0)}$ es necesario simular dos muestras, $x_j^{(k)}$ y $_0 x_j^{(k)}$ (muestra con $\theta = \theta_0$), para así poder calcular $\hat{f}_k(\theta)$ y $\hat{f}_k(\theta_0)$. Si se usa el método de la transformación inversa para simular, podemos usar los mismos números aleatorios para simular ambas muestras. Esta técnica, conocida como Números Aleatorios Comunes (NAC), a menudo sirve para reducir la varianza de estimadores obtenidos por simulación. Cuando se use el mismo número aleatorio para calcular tanto el numerador como el denominador del cociente $\hat{f}(\theta) = \dfrac{\hat{f}_k(\theta)}{\hat{f}_k(\theta_0)}$ se denomina al estimador de $f(\theta)$ resultante $\hat{f}_{NAC}(\theta)$.

## 4.2 Ejemplos y Resultados

Sea $y_{ij}$ la i-ésima observación de la j-ésima muestra con cierta distribución y parámetro $\theta$ y, similarmente, sea $_0 y_{ij}$ la i-ésima observación de la j-ésima muestra con la misma distribución y con parámetro $\theta_0$.

### 4.2.1 Distribución Exponencial

En este caso tenemos que:

$$\hat{f}(\theta) = \frac{\hat{f}_k(\theta)}{\hat{f}_k(\theta_0)} = \frac{\exp\left[\frac{1}{k}\sum_{j=1}^{k} r_j(\theta)\right]}{\exp\left[\frac{1}{k}\sum_{j=1}^{k} r_j(\theta_0)\right]} = \frac{\exp\left[\frac{1}{k}\sum_{j=1}^{k}\ln\left(\frac{\theta^k\left(\sum_{i=1}^{k} y_{ij}\right)^{k+1}}{e^{\theta\sum_{i=1}^{k} y_{ij}} k!}\right)\right]}{\exp\left[\frac{1}{k}\sum_{j=1}^{k}\ln\left(\frac{\theta_0^k\left(\sum_{i=1}^{k} {}_0y_{ij}\right)^{k+1}}{e^{\theta_0\sum_{i=1}^{k} {}_0y_{ij}} k!}\right)\right]} =$$

$$\frac{\exp\left[\ln\prod_{j=1}^{k}\left(\frac{\theta^k\left(\sum_{i=1}^{k} y_{ij}\right)^{k+1}}{e^{\theta\sum_{i=1}^{k} y_{ij}} k!}\right)^{1/k}\right]}{\exp\left[\ln\prod_{j=1}^{k}\left(\frac{\theta_0^k\left(\sum_{i=1}^{k} {}_0y_{ij}\right)^{k+1}}{e^{\theta_0\sum_{i=1}^{k} {}_0y_{ij}} k!}\right)^{1/k}\right]} = \frac{\prod_{j=1}^{k}\left(\frac{\theta^k\left(\sum_{i=1}^{k} y_{ij}\right)^{k+1}}{e^{\theta\sum_{i=1}^{k} y_{ij}} k!}\right)^{1/k}}{\prod_{j=1}^{k}\left(\frac{\theta_0^k\left(\sum_{i=1}^{k} {}_0y_{ij}\right)^{k+1}}{e^{\theta_0\sum_{i=1}^{k} {}_0y_{ij}} k!}\right)^{1/k}} = \left(\frac{\theta}{\theta_0}\right)^k\left(\prod_{j=1}^{k}\left(\frac{\sum_{i=1}^{k} y_{ij}}{\sum_{i=1}^{k} {}_0y_{ij}}\right)^{k+1}\left(\frac{e^{\theta_0\sum_{i=1}^{k} {}_0y_{ij}}}{e^{\theta\sum_{i=1}^{k} y_{ij}}}\right)\right)^{1/k}$$

Usando números aleatorios comunes (NAC) para simular $x_j^{(k)}$ y ${}_0x_j^{(k)}$, se tiene que

${}_0y_{ij} = y_{ij}\left(\dfrac{\theta}{\theta_0}\right)$, por lo que:

$$\hat{f}_{NAC}(\theta) = \left(\frac{\theta}{\theta_0}\right)^k \left(\prod_{j=1}^{k} \left(\frac{\sum_{i=1}^{k} y_{ij}}{\sum_{i=1}^{k} {}_0 y_{ij}}\right)^{k+1} \left(\frac{e^{\theta_0 \sum_{i=1}^{k} {}_0 y_{ij}}}{e^{\theta \sum_{i=1}^{k} y_{ij}}}\right)\right)^{1/k} = \left(\frac{\theta}{\theta_0}\right)^k \left(\prod_{j=1}^{k} \left(\frac{\sum_{i=1}^{k} y_{ij}}{\sum_{i=1}^{k} y_{ij}\left(\frac{\theta}{\theta_0}\right)}\right)^{k+1} \left(\frac{e^{\theta_0 \sum_{i=1}^{k} y_{ij}\left(\frac{\theta}{\theta_0}\right)}}{e^{\theta \sum_{i=1}^{k} y_{ij}}}\right)\right)^{1/k}$$

$$= \left(\frac{\theta}{\theta_0}\right)^k \left(\prod_{j=1}^{k} \left(\frac{\theta_0}{\theta}\right)^{k+1}\right)^{1/k} = \left(\frac{\theta}{\theta_0}\right)^k \left(\frac{\theta_0}{\theta}\right)^{k+1} = \frac{\theta_0}{\theta}$$

En este caso se ve que usando números aleatorios la estimación de $f(\theta)$ es exacta y no depende de la muestra que se simule.

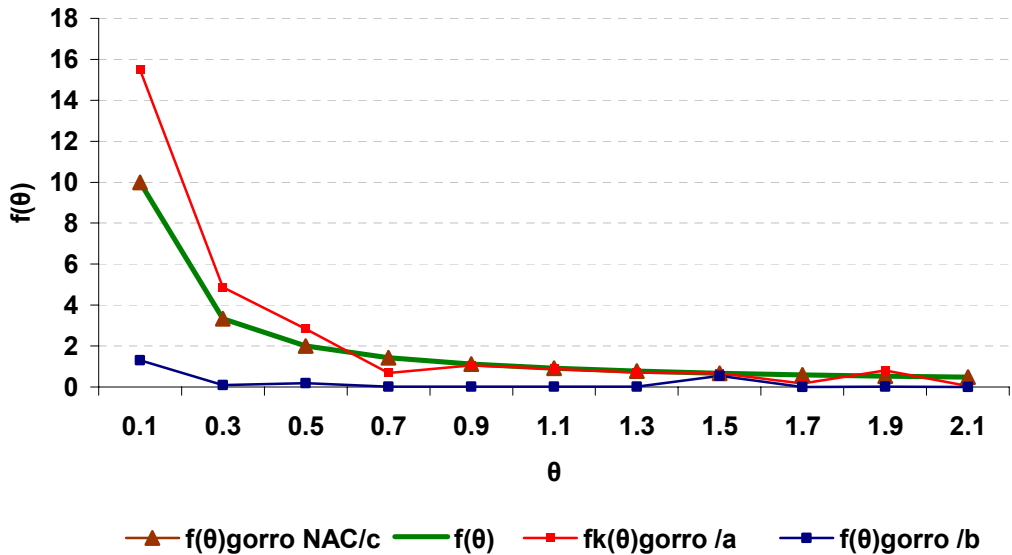

**Figura 4.1: Estimación de** $\frac{f_k(\theta)}{a}$, $\frac{f(\theta)}{b}$ **y** $\frac{f_{NAC}(\theta)}{c}$ **para datos distribuidos** $Exp(\theta)$

La figura 4.1 muestra la estimación de la inicial de referencia para 11 valores de $\theta$ con un tamaño de muestra de de 2 (k=2). La figura 4.1 corrobora que la estimación de $f(\theta)$ con $\frac{\hat{f}_{NAC}(\theta)}{\hat{c}}$ es exacta pues no depende de las muestras que se simulan. Aunque aquí se omite el desarrollo, lo mismo sucede cuando se calcula $\frac{\hat{f}_{NAC}(\theta)}{\hat{c}}$ para datos distribuidos $Unif(0,\theta)$.

Lo que esto significa es que para datos distribuidos $Exp(\theta)$ y $Unif(0,\theta)$, en el calculo de $\dfrac{\hat{f}_{NAC}(\theta)}{\hat{c}}$, da lo mismo simular o no, ya que la expresión no depende de la muestra.

**4.2.2 Distribución Uniforme ($\theta$, $\theta^2$)**

Sea $\quad t_{(k)j} = \max\limits_{i=1..k}\{y_{ij}\}, t_{(1)j} = \min\limits_{i=1..k}\{y_{ij}\}, {}_0t_{(k)j} = \max\limits_{i=1..k}\{{}_0y_{ij}\}, {}_0t_{(1)j} = \min\limits_{i=1..k}\{{}_0y_{ij}\}$,

entonces:

$$\hat{f}(\theta) = \frac{\hat{f}_k(\theta)}{\hat{f}_k(\theta_0)} = \frac{\exp\left[\dfrac{1}{k}\sum_{j=1}^{k}r_j(\theta)\right]}{\exp\left[\dfrac{1}{k}\sum_{j=1}^{k}r_j(\theta_0)\right]} = \frac{\exp\left[\dfrac{1}{k}\sum_{j=1}^{k}\ln\left(\dfrac{1}{\theta^k(\theta-1)^k\left[\int_{\sqrt{t_{(k)j}}}^{t_{(1)j}}\theta^{-k}(\theta-1)^{-k}d\theta\right]}\right)\right]}{\exp\left[\dfrac{1}{k}\sum_{j=1}^{k}\ln\left(\dfrac{1}{\theta_0^k(\theta_0-1)^k\left[\int_{\sqrt{{}_0t_{(k)j}}}^{{}_0t_{(1)j}}\theta^{-k}(\theta-1)^{-k}d\theta\right]}\right)\right]} =$$

$$\frac{\exp\left[\ln\left(\prod_{j=1}^{k}\dfrac{1}{\theta^k(\theta-1)^k\left[\int_{\sqrt{t_{(k)j}}}^{t_{(1)j}}\theta^{-k}(\theta-1)^{-k}d\theta\right]}\right)^{1/k}\right]}{\exp\left[\ln\left(\prod_{j=1}^{k}\dfrac{1}{\theta_0^k(\theta_0-1)^k\left[\int_{\sqrt{{}_0t_{(k)j}}}^{{}_0t_{(1)j}}\theta^{-k}(\theta-1)^{-k}d\theta\right]}\right)^{1/k}\right]} = \frac{\left(\prod_{j=1}^{k}\dfrac{1}{\theta^k(\theta-1)^k\left[\int_{\sqrt{t_{(k)j}}}^{t_{(1)j}}\theta^{-k}(\theta-1)^{-k}d\theta\right]}\right)^{1/k}}{\left(\prod_{j=1}^{k}\dfrac{1}{\theta_0^k(\theta_0-1)^k\left[\int_{\sqrt{{}_0t_{(k)j}}}^{{}_0t_{(1)j}}\theta^{-k}(\theta-1)^{-k}d\theta\right]}\right)^{1/k}}$$

$$= \left(\dfrac{\theta_0(\theta_0-1)}{\theta(\theta-1)}\right)^k\left(\prod_{j=1}^{k}\left(\dfrac{\int_{\sqrt{{}_0t_{(k)j}}}^{{}_0t_{(1)j}}\theta^{-k}(\theta-1)^{-k}d\theta}{\int_{\sqrt{t_{(k)j}}}^{t_{(1)j}}\theta^{-k}(\theta-1)^{-k}d\theta}\right)^{k-1}\right)^{1/k}$$

Usando números aleatorios comunes se tiene que ${}_0t_{(1)j} = \dfrac{(t_{(1)j}-\theta)\theta_0(\theta_0-1)}{\theta(\theta-1)} + \theta_0$ y

${}_0t_{(k)j} = \dfrac{(t_{(k)j}-\theta)\theta_0(\theta_0-1)}{\theta(\theta-1)} + \theta_0$ por lo que:

$$\hat{f}_{NAC}(\theta) = \left(\frac{\theta_0(\theta_0-1)}{\theta(\theta-1)}\right)^k \left(\prod_{j=1}^{k} \left(\frac{\int_b^a \theta^{-k}(\theta-1)^{-k} d\theta}{\int_{\sqrt{t_{(k)j}}}^{t_{(1)j}} \theta^{-k}(\theta-1)^{-k} d\theta}\right)^{k-1}\right)^{1/k}$$ con

$$a = \frac{(t_{(1)j}-\theta)\theta_0(\theta_0-1)}{\theta(\theta-1)} + \theta_0 \ \text{y} \ b = \sqrt{\frac{(t_{(k)j}-\theta)\theta_0(\theta_0-1)}{\theta(\theta-1)}} + \theta_0.$$

En este caso el estimador $\hat{f}_{NAC}(\theta)$, no parece poder reducirse a una expresión sencilla que no dependa de la muestra.

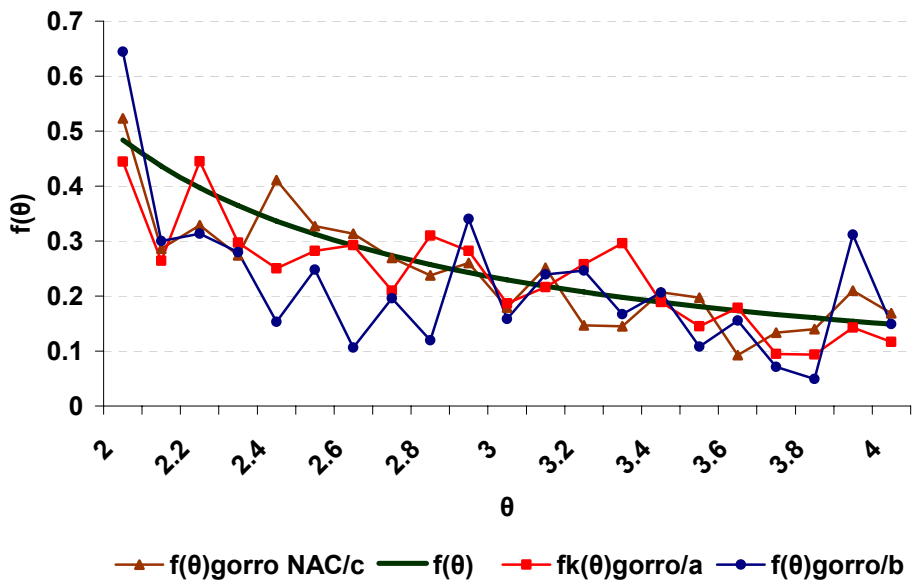

**Figura 4.2: Estimación de** $\frac{f_k(\theta)}{a}$ **,** $\frac{f(\theta)}{b}$ **y** $\frac{f_{NAC}(\theta)}{c}$ **para datos distribuidos** $Unif(\theta, \theta^2)$

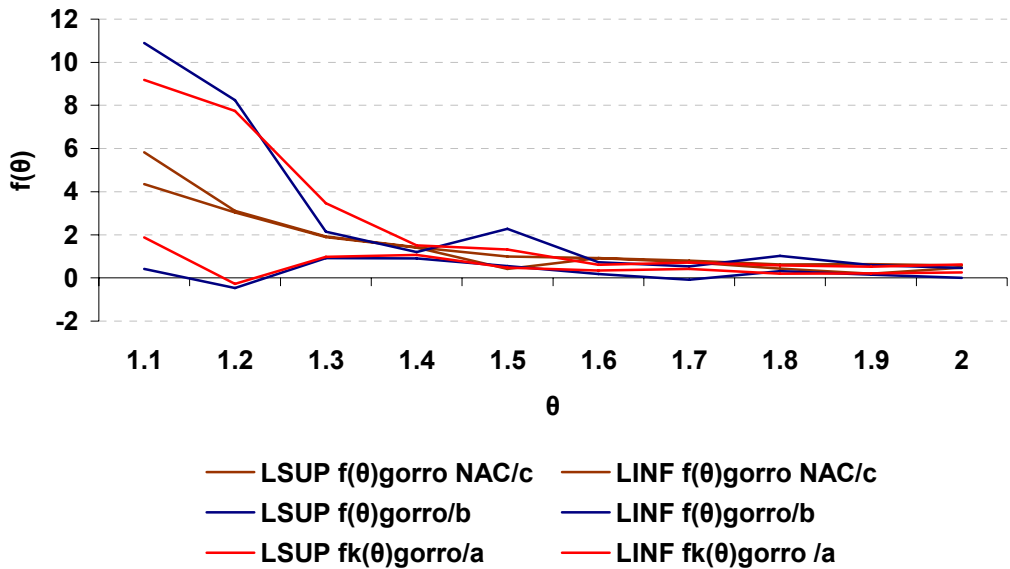

**Figura 4.3: Límites inferior y superior de** $\dfrac{\hat{f}_k(\theta)}{\hat{a}}$, $\dfrac{\hat{f}(\theta)}{\hat{b}}$ **y** $\dfrac{\hat{f}_{NAC}(\theta)}{\hat{c}}$ **para datos distribuidos** $Unif(\theta, \theta^2)$

La figura 4.2 muestra la estimación de la inicial de referencia para 21 valores de $\theta$ con un tamaño de muestra de de 2 (k=2). En este caso la estimación de $f(\theta)$ con $\dfrac{\hat{f}_{NAC}(\theta)}{\hat{c}}$ no es exacta. La figura 4.3 muestra los límites inferior y superior para los estimadores $\dfrac{\hat{f}_k(\theta)}{\hat{a}}$, $\dfrac{\hat{f}(\theta)}{\hat{b}}$ y $\dfrac{\hat{f}_{NAC}(\theta)}{\hat{c}}$ para 10 distintos valores de $\theta$ y con un tamaño de muestra de 5 y valor de alfa igual a 0.1. Parece que, por lo menos para valores de $\theta$ menores a 1.4, de los tres estimadores el que tiene intervalos más angostas y por lo tanto es más preciso es $\dfrac{\hat{f}_{NAC}(\theta)}{\hat{c}}$ usando números aleatorios comunes.

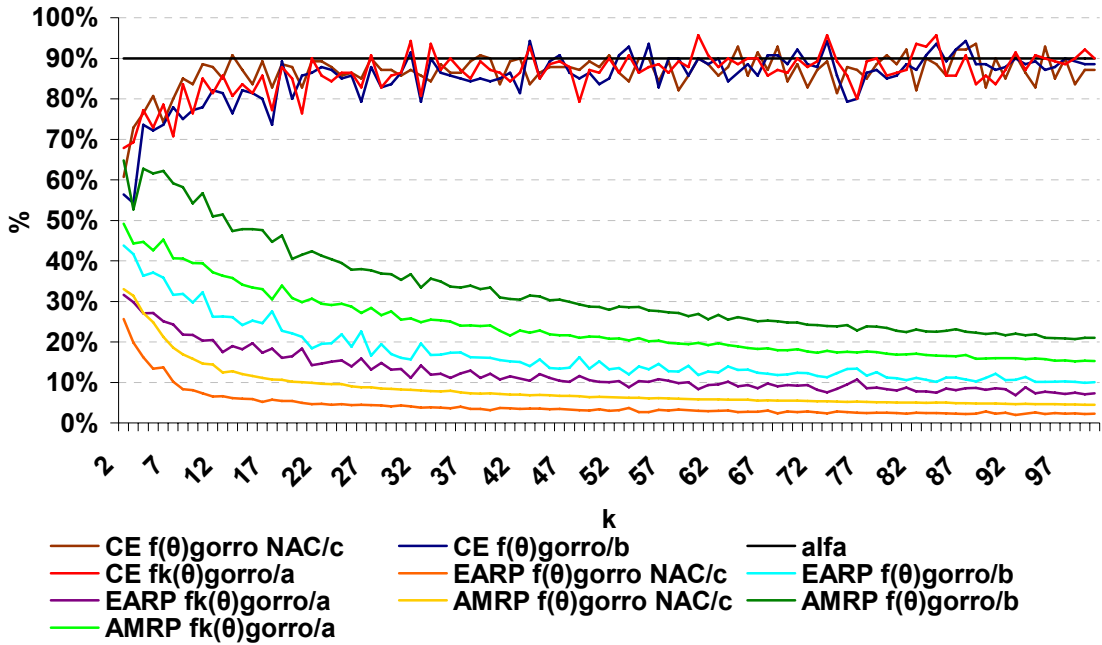

**Figura 4.4: CE, AMRP y EARP de** $\dfrac{\hat{f}_k(\theta)}{\hat{a}}$ , $\dfrac{\hat{f}(\theta)}{\hat{b}}$ **y** $\dfrac{\hat{f}_{NAC}(\theta)}{\hat{c}}$ **para datos distribuidos** $Unif(\theta, \theta^2)$

La figura 4.4 muestra la cobertura empírica, ancho medio relativo promedio y error absoluto relativo promedio de los estimadores $\dfrac{\hat{f}_k(\theta)}{\hat{a}}$ , $\dfrac{\hat{f}(\theta)}{\hat{b}}$ y $\dfrac{\hat{f}_{NAC}(\theta)}{\hat{c}}$ conforme k crece. Para cada valor de k se simularon 140 valores de $\theta$. Los límites inferior y superior se construyeron con un valor de alfa igual a 0.1. Se muestra que para tamaños de muestra mayores a 35 los tres estimadores logran coberturas empíricas similares $1-\alpha$. El AMRP del estimador $\dfrac{\hat{f}_{NAC}(\theta)}{\hat{c}}$ es siempre mucho menor al de $\dfrac{\hat{f}_k(\theta)}{\hat{a}}$ y $\dfrac{\hat{f}(\theta)}{\hat{b}}$ y como las cobertura empírica de $\dfrac{\hat{f}_{NAC}(\theta)}{\hat{c}}$ es mayor a la de $\dfrac{\hat{f}_k(\theta)}{\hat{a}}$ y $\dfrac{\hat{f}(\theta)}{\hat{b}}$ para tamaños de muestra menores a 20 y muy similar para tamaños de muestra mayores el EARP de $\dfrac{\hat{f}_{NAC}(\theta)}{\hat{c}}$ es menor al de $\dfrac{\hat{f}_k(\theta)}{\hat{a}}$ y $\frac{\hat{f}(\theta)}{\hat{b}}$. Visto de otra forma se necesita un tamaño de muestra de alrededor de 10 para obtener un EARP de 10% con $\frac{\hat{f}_{NAC}(\theta)}{\hat{c}}$ mientras que con $\frac{\hat{f}_k(\theta)}{\hat{a}}$ y $\frac{\hat{f}(\theta)}{\hat{b}}$ se necesitan tamaños de muestra de alrededor de 60 y 100 respectivamente para obtener un EARP similar.

### 4.2.3 Distribución Triangular

Sea $t_{(r)j}$ la r-ésima estadística de orden de la j-ésima muestra distribuida $Trian(0,\theta,1)$ y sea $_0t_{(r)j}$ la r-ésima estadística de orden de la j-ésima muestra distribuida $Trian(0,\theta_0,1)$, entonces:

$$c_j = 2^k \left[ \left(\prod_{i=1}^{k} 1-t_{(i)j}\right)\left(\frac{(1-t_{(1)j})^{1-k}-1}{k-1}\right) + \left(\prod_{i=1}^{1} t_{(i)j}\right)\left(\prod_{i=2}^{k} 1-t_{(i)j}\right)\int_{t_{(1)j}}^{t_{(2)j}} \left(\frac{1}{\theta}\right)^1 \left(\frac{1}{1-\theta}\right)^{k-1} d\theta \right. \\ +\ldots+ \left(\prod_{i=1}^{q} t_{(i)j}\right)\left(\prod_{i=q+1}^{k} 1-t_{(i)j}\right)\int_{t_{(q)j}}^{t_{(q+1)j}} \left(\frac{1}{\theta}\right)^q \left(\frac{1}{1-\theta}\right)^{k-(q+1)} d\theta +\ldots+ \\ \left. \left(\prod_{i=1}^{k-1} t_{(i)j}\right)\left(\prod_{i=k}^{k} 1-t_{(i)j}\right)\int_{t_{(k-1)j}}^{t_{(k)j}} \left(\frac{1}{\theta}\right)^{k-1} \left(\frac{1}{1-\theta}\right)^{k-(k-1)} d\theta + \left(\prod_{i=1}^{k} t_{(i)j}\right)\left(\frac{(t_{(k)j})^{1-k}-1}{k-1}\right) \right]$$

$$= 2^k g(k, t_{(1)j}, t_{(2)j}, \ldots t_{(k)j})$$

De manera similar $_0c_j = 2^k g(k, {_0t_{(1)j}}, {_0t_{(2)j}}, \ldots, {_0t_{(k)j}})$, por lo que:

$$r_j(\theta) = \ln\left(\frac{2^k\left(\prod_{i=1}^{r}t_{(i)j}\right)\left(\prod_{i=r+1}^{k}1-t_{(i)j}\right)\left(\frac{1}{\theta}\right)^r\left(\frac{1}{1-\theta}\right)^{k-r}}{2^k g(k, t_{(1)j}, t_{(2)j}, \ldots, t_{(k)j})}\right) = \ln\left(\frac{\left(\prod_{i=1}^{r}t_{(i)j}\right)\left(\prod_{i=r+1}^{k}1-t_{(i)j}\right)}{g(k, t_{(1)j}, t_{(2)j}, \ldots, t_{(k)j})\theta^r(1-\theta)^{k-r}}\right)$$

Con $t_{(r)j} < \theta < t_{(r+1)j}$.

De ahí que,

$$\hat{f}(\theta) = \frac{\hat{f}_k(\theta)}{\hat{f}_k(\theta_0)} = \frac{\exp\left[\frac{1}{k}\sum_{j=1}^{k}r_j(\theta)\right]}{\exp\left[\frac{1}{k}\sum_{j=1}^{k}r_j(\theta_0)\right]} = \frac{\exp\left[\frac{1}{k}\sum_{j=1}^{k}\ln\left(\frac{\left(\prod_{i=1}^{r}t_{(i)j}\right)\left(\prod_{i=r+1}^{k}1-t_{(i)j}\right)}{g(k, t_{(1)j}, t_{(2)j}, \ldots, t_{(k)j})\theta^r(1-\theta)^{k-r}}\right)\right]}{\exp\left[\frac{1}{k}\sum_{j=1}^{k}\ln\left(\frac{\left(\prod_{i=1}^{r_0}{}_0t_{(i)j}\right)\left(\prod_{i=r_0+1}^{k}1-{}_0t_{(i)j}\right)}{g(k, {}_0t_{(1)j}, {}_0t_{(2)j}, \ldots, {}_0t_{(k)j})\theta_0^{r_0}(1-\theta_0)^{k-r_0}}\right)\right]} =$$

$$\frac{\exp\left[\ln\left(\prod_{j=1}^{k}\frac{\left(\prod_{i=1}^{r}t_{(i)j}\right)\left(\prod_{i=r+1}^{k}1-t_{(i)j}\right)}{g(k, t_{(1)j}, t_{(2)j}, \ldots, t_{(k)j})\theta^r(1-\theta)^{k-r}}\right)^{1/k}\right]}{\exp\left[\ln\left(\prod_{j=1}^{k}\frac{\left(\prod_{i=1}^{r_0}{}_0t_{(i)j}\right)\left(\prod_{i=r_0+1}^{k}1-{}_0t_{(i)j}\right)}{g(k, {}_0t_{(1)j}, {}_0t_{(2)j}, \ldots, {}_0t_{(k)j})\theta_0^{r_0}(1-\theta_0)^{k-r_0}}\right)^{1/k}\right]} = \frac{\left(\prod_{j=1}^{k}\frac{\left(\prod_{i=1}^{r}t_{(i)j}\right)\left(\prod_{i=r+1}^{k}1-t_{(i)j}\right)}{g(k, t_{(1)j}, t_{(2)j}, \ldots, t_{(k)j})\theta^r(1-\theta)^{k-r}}\right)^{1/k}}{\left(\prod_{j=1}^{k}\frac{\left(\prod_{i=1}^{r_0}{}_0t_{(i)j}\right)\left(\prod_{i=r_0+1}^{k}1-{}_0t_{(i)j}\right)}{g(k, {}_0t_{(1)j}, {}_0t_{(2)j}, \ldots, {}_0t_{(k)j})\theta_0^{r_0}(1-\theta_0)^{k-r_0}}\right)^{1/k}}$$

$$= \frac{\theta_0^{r_0}(1-\theta_0)^{k-r_0}}{\theta^r(1-\theta)^{k-r}}\left[\prod_{j=1}^{k}\frac{g(k, {}_0t_{(1)j}, {}_0t_{(2)j}, \ldots, {}_0t_{(k)j})\left(\prod_{i=1}^{r}t_{(i)j}\right)\left(\prod_{i=r+1}^{k}1-t_{(i)j}\right)}{g(k, t_{(1)j}, t_{(2)j}, \ldots, t_{(k)j})\left(\prod_{i=1}^{r_0}{}_0t_{(i)j}\right)\left(\prod_{i=r_0+1}^{k}1-{}_0t_{(i)j}\right)}\right]^{1/k}$$

Usando números aleatorios comunes se tiene lo siguiente:

$${}_0t_{(a)j} = t_{(a)j}\sqrt{\frac{\theta_0}{\theta}}$ si $t_{(a)j} \leq \theta, {}_0t_{(a)j} \leq \theta_0$



$$_0t_{(a)j} = \sqrt{\theta_0\left(1 - \frac{(1-t_{(a)j})^2}{1-\theta}\right)} \text{ si } t_{(a)j} > \theta, _0t_{(a)j} \leq \theta_0$$

$$_0t_{(a)j} = 1 - \sqrt{(1-\theta_0)\left(1 - \frac{t_{(a)j}^2}{\theta}\right)} \text{ si } t_{(a)j} \leq \theta, _0t_{(a)j} > \theta_0$$

$$_0t_{(a)j} = 1 - \left[(1-t_{(a)j})\sqrt{\frac{1-\theta_0}{1-\theta}}\right] \text{ si } t_{(a)j} > \theta, _0t_{(a)j} > \theta_0 \text{ para } a \in \{1,...,k\}$$

Con esto y desarrollando la expresión que ya se tiene para $\hat{f}(\theta)$ se llega a que:

Si $\theta \leq \theta_0 \Rightarrow r \leq r_0$ y

$$\hat{f}_{NAC}(\theta) = \frac{\theta_0^{r_0/2}(1-\theta_0)^{(k-r_0)/2}}{\theta^{r/2}(1-\theta)^{(k-r)-(k-r_0)/2}} \left[\prod_{j=1}^{k} \frac{g(k,x_0^{(j)})\left(\prod_{i=r+1}^{r_0} 1 - t_{(i)j}\right)}{g(k,x^{(j)})\left(\prod_{i=r+1}^{r_0} \sqrt{1 - \frac{(1-t_{(i)j})^2}{1-\theta}}\right)}\right]^{1/k}$$

Si $\theta > \theta_0 \Rightarrow r \geq r_0$ y

$$\hat{f}_{NAC}(\theta) = \frac{\theta_0^{r_0/2}(1-\theta_0)^{(k-r_0)/2}}{\theta^{r-r_0/2}(1-\theta)^{(k-r)/2}} \left[\prod_{j=1}^{k} \frac{g(k,x_0^{(j)})\left(\prod_{i=r_0+1}^{r} t_{(i)j}\right)}{g(k,x^{(j)})\left(\prod_{i=r_0+1}^{r} \sqrt{\left[1 - \frac{t_{(i)j}^2}{\theta}\right]}\right)}\right]^{1/k}$$

Como en el caso anterior el estimador $\hat{f}_{NAC}(\theta)$ no parece poder reducirse a una expresión sencilla que no dependa de la muestra.

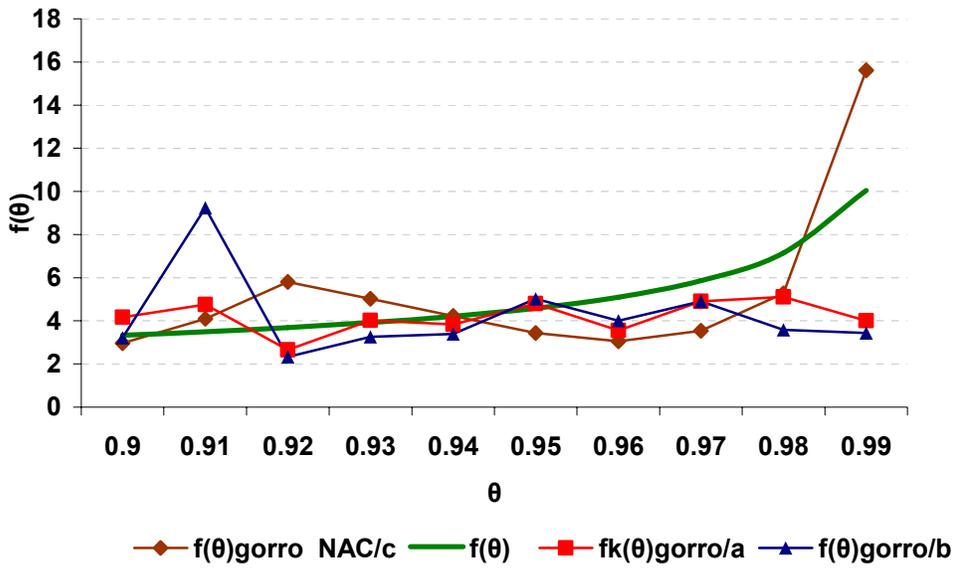

**Figura 4.5:** Estimación de $\dfrac{f_k(\theta)}{a}$, $\dfrac{f(\theta)}{b}$ y $\dfrac{f_{NAC}(\theta)}{c}$ para datos distribuidos $Trian(0,\theta,1)$

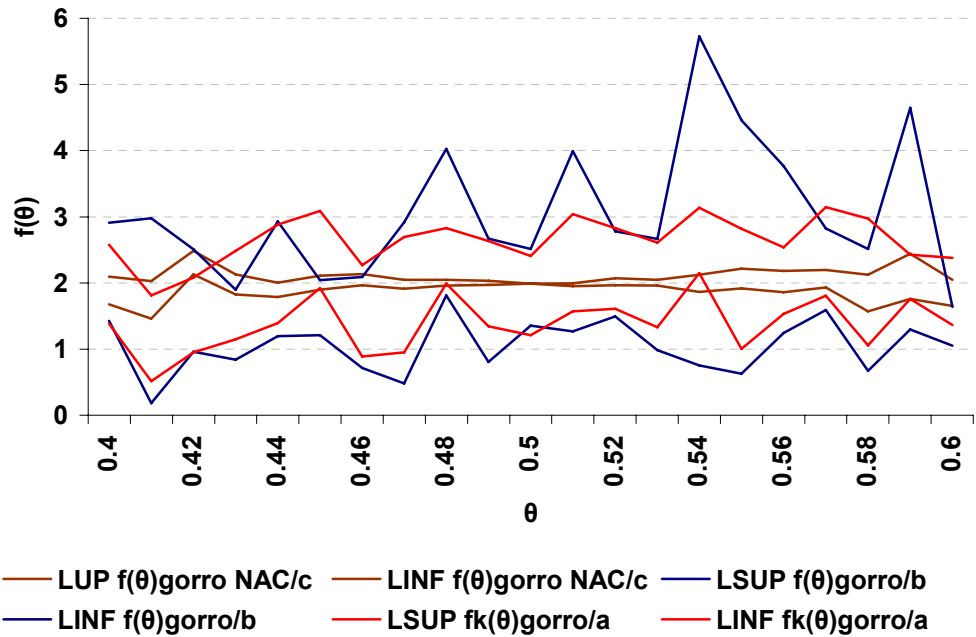

**Figura 4.6:** Límites inferior y superior de $\dfrac{\hat{f}_k(\theta)}{\hat{a}}$, $\dfrac{\hat{f}(\theta)}{\hat{b}}$ y $\dfrac{\hat{f}_{NAC}(\theta)}{\hat{c}}$ para datos distribuidos $Trian(0,\theta,1)$

La figura 4.5 muestra la estimación de $f(\theta)$ para 10 valores de $\theta$ y un tamaño de muestra de 10 (k=10). En este caso no hay una clara diferencia en la precisión de los estimadores. Se tendrá que estudiar qué sucede con el error absoluto relativo promedio y el ancho relativo promedio conforme k crece para determinar si existe alguna diferencia. La figura 4.6 muestra la los límites inferior y superior de tres estimadores de $f(\theta)$ para 21 distintos valores de $\theta$ y un tamaño de muestra de 10 (k=10) y un valor de alfa igual a 0.08. Parece que $\dfrac{\hat{f}_{NAC}(\theta)}{\hat{c}}$ es más preciso que $\dfrac{\hat{f}_k(\theta)}{\hat{a}}$ y $\dfrac{\hat{f}(\theta)}{\hat{b}}$ puesto que sus intervalos son más angostos.

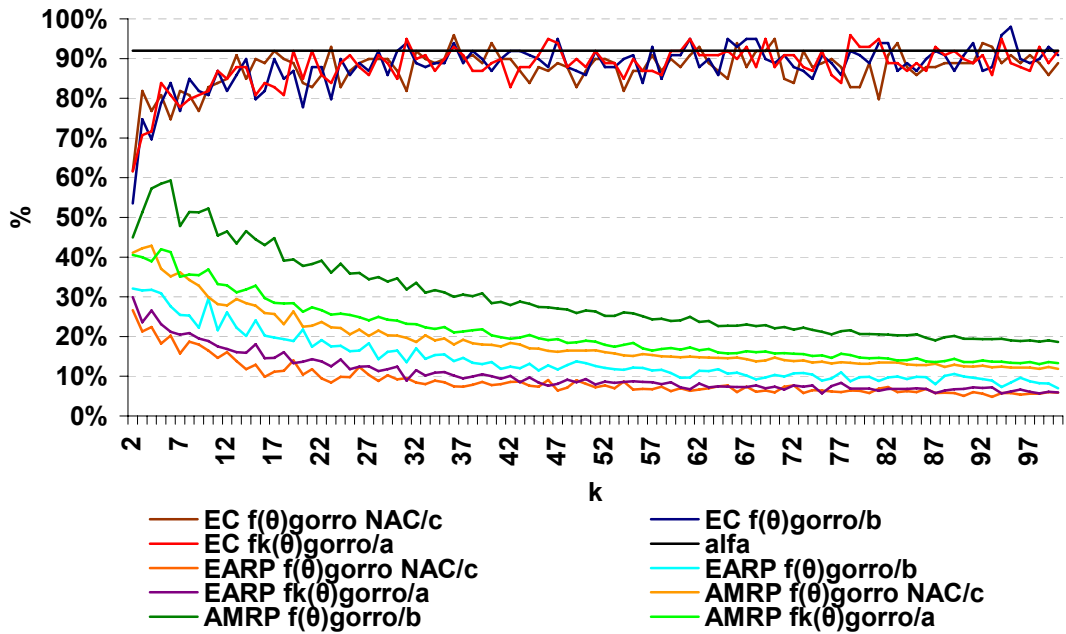

**Figura 4.7: CE, AMRP y EARP de** $\dfrac{\hat{f}_k(\theta)}{\hat{a}}$ **,** $\dfrac{\hat{f}(\theta)}{\hat{b}}$ **y** $\dfrac{\hat{f}_{NAC}(\theta)}{\hat{c}}$ **para datos distribuidos** $Trian(0,\theta,1)$

La figura 4.7 muestra la cobertura empírica, ancho medio relativo promedio y error absoluto relativo promedio de los estimadores $\dfrac{\hat{f}_k(\theta)}{\hat{a}}$ , $\dfrac{\hat{f}(\theta)}{\hat{b}}$ y $\dfrac{\hat{f}_{NAC}(\theta)}{\hat{c}}$ conforme k crece. Para cada valor de k se simularon 99 valores de $\theta$. Los límites inferior y superior se construyeron con un valor de alfa igual a 0.08. Se muestra que para tamaños de muestra mayores a 35 los tres estimadores logran coberturas empíricas similares $1-\alpha$. El AMRP del estimador $\dfrac{\hat{f}_{NAC}(\theta)}{\hat{c}}$ es, en general, un poco menor al de $\dfrac{\hat{f}_k(\theta)}{\hat{a}}$ y como las cobertura empírica de $\dfrac{\hat{f}_{NAC}(\theta)}{\hat{c}}$ es muy similar a la de $\dfrac{\hat{f}_k(\theta)}{\hat{a}}$ el EARP de $\dfrac{\hat{f}_{NAC}(\theta)}{\hat{c}}$ es, en general, menor al de $\dfrac{\hat{f}_k(\theta)}{\hat{a}}$. Sin embargo, en este caso, para tamaños de muestra superiores a 50 el EARP de ambos estimadores es muy similar. Visto de otra forma se necesita un tamaño de muestra de alrededor de 25 para obtener un EARP de 10% con $\dfrac{\hat{f}_{NAC}(\theta)}{\hat{c}}$ mientras que con $\dfrac{\hat{f}_k(\theta)}{\hat{a}}$ se necesitan tamaños de muestra de alrededor de 40 para obtener un EARP similar.

### 4.3 Conclusiones

Para los tres ejemplos tratados el estimador $\dfrac{\hat{f}_{NAC}(\theta)}{\hat{c}}$ fue el más preciso de los estimadores de $f(\theta)$. Para datos distribuidos $Exp(\theta)$ la estimación es exacta puesto que la expresión $\hat{f}_{NAC}(\theta) = \dfrac{\hat{f}_k(\theta)}{\hat{f}_k(\theta_0)}$ se reduce a una que no depende de las muestras simuladas. En el caso de datos distribuidos $Unif(\theta, \theta^2)$ o $Trian(0, \theta, 1)$ esta expresión no parece poderse reducir a una expresión que no dependa de la muestra. Aún así, en estos casos, el estimador $\dfrac{\hat{f}_{NAC}(\theta)}{\hat{c}}$ es más preciso que $\dfrac{\hat{f}_k(\theta)}{\hat{a}}$ y $\dfrac{\hat{f}(\theta)}{\hat{b}}$ puesto que para un mismo valor de $\theta$ y $\alpha$ el ancho medio es menor y como la cobertura empírica es similar para ambos, el EARP es menor. Sin embargo en el caso de datos distribuidos $Trian(0,\theta,1)$ la ventaja en la precisión del estimador $\dfrac{\hat{f}_{NAC}(\theta)}{\hat{c}}$ solo es significativa para tamaños de muestra menores a 50.

## 5. CONCLUSIONES

Los dos resultados más importantes de esta investigación son:

1. La derivación de la Varianza del estimador de Bernardo y la construcción, con ésta, de intervalos de probabilidad que permiten relacionar el tamaño de muestra que se usa en el método de simulación con el error de estimación del mismo.

2. Usando números aleatorios comunes, para datos distribuidos $Unif(\theta, \theta^2)$ o $Trian(0, \theta, 1)$ se observó que en el cálculo de $\hat{f}_{NAC}(\theta)$ se reduce considerablemente el error de estimación medido por el EARP. Inclusive se vio que para el caso de datos distribuidos $Exp(\theta)$, el error es nulo ya que el estimador resulta no depender de la muestra. Este hallazgo justifica el uso del estimador $\hat{f}_{NAC}(\theta)$ sobre $\hat{f}_k(\theta)$ ya que no es posible utilizar números aleatorios comunes en el cálculo de $\hat{f}_k(\theta)$.

La realización de este trabajo fue una experiencia muy enriquecedora para mi sobre todo porque fue la primera vez que lleve a cabo un trabajo largo, cuya complejidad requería de una estructuración del problema en distintos sub-problemas para poder atacarlo de forma eficaz. Además, no habiendo estudiado la estadística Bayesiana previo a la realización de este trabajo, me resulto muy atractiva la forma de atacar los problemas de decisión en ambiente de incertidumbre de la teoría Bayesiana puesto que provee un marco de referencia básico con el cual poder plantear de inicio dichos problemas y sobre el cual puede uno inscribir y relacionar los distintos métodos estadísticos, a menudo compatibles, de la estadística frecuentista así como la teoría de probabilidad, formándose uno un mapa mental integral de cómo atacar problemas de decisión en contextos en los cuales existe incertidumbre, ya sea porque se tiene un fenómeno que presenta variabilidad o porque simplemente se tiene información limitada sobre el fenómeno a modelar. Por otro lado el análisis de las iniciales de referencia requiere de sólidas bases en los campos de análisis matemático y teoría de medida y mi falta de profundización en estos temas limito mi entendimiento de la importancia de las iniciales de referencia en el contexto de las funciones iniciales en general.

# 6. BIBLIOGRAFÍA

Berger, James, José M. Bernardo y Dongchu Sun. "The formal definition of reference priors". *Annals of Statistics* V.37, N.2 (2009): pp.905-938.

Bernardo, José M. "Reference Analysis". *Handbook of Statistics.* V.25 (2005): pp. 17-90.

Bernardo, José M. y Adrian F. M. Smith. (2000). *Bayesian Theory.* Chichester: John Wiley.

Billingsley, Patrick (1986). *Convergence of Probability Measures*. New York: John Wiley.

Chung Kai Lai (1974). *A Course in Probability Theory*, 2/e. San Diego: Academic Press.

Ethier, Stuart N. y Thomas G Kurtz, (1968). *Markov Processes, Characterization and Convergence*. New York: John Wiley.

Mood, Alexander, Franklin A. Graybill y Duane C. Boes (1974). *Introduction to the Theory of Statistics*, 3/e. New York: McGrawHill.

Muñoz, David F. y Peter W Glynn, "A batch means methodology for estimation of a nonlinear function of a steady-state mean". *Management Science.* V.43, N. 8 (1997): pp 1121-1135.

Serfling Robert J.(1980). *Approximation Theorems of Mathematical Statistics*. New York: John Wiley.